\documentclass[12pt]{article}
\pdfoutput=1

\usepackage{amsmath}
\usepackage{amsfonts}
\usepackage{amssymb}
\usepackage{dsfont}

\usepackage{tikz}
\usetikzlibrary{decorations.text}
\usetikzlibrary{decorations.pathreplacing}
\usetikzlibrary{calc}

\usepackage{booktabs}

\usepackage{hyperref}
\hypersetup{colorlinks=true, linkcolor=red, citecolor=blue, linktoc=page}

\usepackage{color}
\usepackage{graphicx}

\usepackage{sectsty}
\sectionfont{\large}
\renewcommand{\thefootnote}{\fnsymbol{footnote}}
\newcommand{\starttext}{
\setcounter{footnote}{0}
\renewcommand{\thefootnote}{\arabic{footnote}}}
\newcommand{\no}{\nonumber}

\newcommand{\ZZ}{\mathds{Z}}

\numberwithin{equation}{section}

\newcommand{\cA}{\mathcal A}
\newcommand{\cB}{\mathcal B}
\newcommand{\cC}{\mathcal C}
\newcommand{\cG}{\mathcal G}

\DeclareMathOperator{\Vol}{Vol}
\DeclareMathOperator{\vol}{vol}

\def\bea{\begin{eqnarray}}
\def\eea{\end{eqnarray}}
\def\be{\begin{equation}}
\def\ee{\end{equation}}
\newcommand\cZ{\mathcal{Z}}
\newcommand\cF{\mathcal{F}}
\newcommand\cW{\mathcal{W}}
\newcommand\cI{\mathcal{I}}
\newcommand\cO{\mathcal{O}}
\newcommand\diff{\mathrm{d}}
\newcommand{\ii}{\mathrm{i}}
\newcommand{\nn}{\nonumber}
\usepackage{multirow}
\def\ie{\begin{equation}\begin{aligned}}
\def\fe{\end{aligned}\end{equation}}
\usepackage{subfig}
\newcommand\cQ{\mathcal{Q}}
\usepackage{float}
\usepackage{cite}

\begin{document}
\setlength{\baselineskip}{16pt}

\starttext
\setcounter{footnote}{0}

\begin{flushright}
CALT-TH 2018-022\\
\today
\end{flushright}

\vskip 1in

\begin{center}

{\Large \bf  Precision test of $\mathbf{AdS_6/CFT_5}$ in Type IIB}

\vskip 0.4in

{\large   Martin Fluder$^a$, Christoph F.~Uhlemann$^b$}

\vskip 0.2in

{${}^a$ Walter Burke Institute for Theoretical Physics \\ California Institute of Technology,
Pasadena, CA 91125, USA}

\vskip 0.2in

{\sl $^b$ Mani L. Bhaumik Institute for Theoretical Physics} \\
{\sl Department of Physics and Astronomy }\\
{\sl University of California, Los Angeles, CA 90095, USA} 

\vskip 0.2in

{\tt fluder@caltech.edu, uhlemann@physics.ucla.edu}

\vskip 0.8in

\end{center}
 
\begin{abstract}
\setlength{\baselineskip}{16pt}

Large classes of warped AdS$_6$ solutions were constructed recently in Type IIB supergravity, and identified as holographic duals for five-dimensional superconformal field theories realized by $(p,q)$ five-brane webs. We confront holographic results for the five sphere partition functions obtained from these solutions with computations for the putative dual field theories. We obtain the sphere partition functions and conformal central charges in gauge theory deformations of the superconformal field theories numerically using supersymmetric localization, and extrapolate the results to the conformal fixed points. In the appropriate large $N$ limits, the results match precisely to the supergravity computations, providing strong support for the proposed dualities.

\noindent 

\end{abstract}

\setcounter{equation}{0}
\setcounter{footnote}{0}

\newpage

\tableofcontents

\section{Introduction}

Five-dimensional superconformal field theories are interesting for a variety of reasons. On the one hand, defining interacting quantum field theories in dimensions greater than four is challenging, as conventional gauge theories are perturbatively non-renormalizable. On the other hand, a large body of evidence suggests that interacting superconformal field theories in five and six dimensions exist, and have interesting relations to field theories in lower dimensions~\cite{Seiberg:1996bd,Morrison:1996xf,Intriligator:1997pq,Aharony:1997ju,Aharony:1997bh}. Many of the five-dimensional superconformal field theories have relevant deformations that flow to conventional, although non-renormalizable, five-dimensional gauge theories in the infrared, rendering these gauge theories ``asymptotically safe''. The ultraviolet fixed points themselves, however, do not have supersymmetric marginal deformations (no dimensionless coupling constants) and no conventional Lagrangian description.

In the absence of a conventional Lagrangian description, indirect tools such as AdS/CFT dualities are particularly useful for quantitative studies of the five-dimensional superconformal field theories. A prerequisite for this approach is the availability of AdS$_6$ supergravity solutions and a precise identification of such solutions with dual field theories. A warped AdS$_6$ solution in massive Type IIA supergravity has been known for some time~\cite{Brandhuber:1999np,Ferrara:1998gv,Bergman:2012kr}, and has been studied and extensively checked~\cite{Ferrara:1998gv,Brandhuber:1999np,Bergman:2012kr,Jafferis:2012iv,Bergman:2012qh,Passias:2012vp,Assel:2012nf,Bergman:2013koa,Alday:2014rxa,Alday:2014bta,Alday:2014fsa,Hama:2014iea,Alday:2015jsa}. More recently, large classes of warped AdS$_6$ solutions were constructed in Type IIB supergravity~\cite{DHoker:2016ujz,DHoker:2016ysh,DHoker:2017mds,DHoker:2017zwj}. These solutions provide holographic duals for large classes of five-dimensional superconformal field theories, and there are compelling arguments for their identification with $(p,q)$ five-brane webs in Type IIB string theory. This identification suggests a concrete map between supergravity solutions and field theories. Several aspects of the dualities have since been studied holographically~\cite{Gutperle:2017tjo,Gutperle:2018vdd,Gutperle:2018wuk,Kaidi:2017bmd}, and the spectrum of certain large-scaling-dimension operators for various superconformal field theories has been matched to supergravity in~\cite{Bergman:2018}.

The aim of this work is to provide a decisive test of the aforementioned identification of Type IIB supergravity solutions with five-dimensional superconformal field theories at the level of the partition functions. We will obtain the partition functions of the superconformal field theories on the five-sphere, ${\rm S}^5$, as well as the conformal central charges $C_T$, which are related to the partition functions on squashed five-spheres. The strategy will be to study gauge theory deformations of the superconformal field theories, and use supersymmetric localization~\cite{Pestun:2007rz,Kallen:2012va,Kim:2012ava,Imamura:2012bm,Lockhart:2012vp,Imamura:2012xg} to compute the partition functions at strong coupling. The results will then be extrapolated to the ultraviolet fixed-point theories. We will study two specific examples: The $T_N$ theories~\cite{Benini:2009gi,Bao:2013pwa}, which reduce to isolated, intrinsically strongly-coupled superconformal field theories in four dimensions upon compactification, and the theories realized on intersections of $N$ D5 and $M$ NS5-branes, which we will refer to as the $\#_{N,M}$ theories, studied originally in~\cite{Aharony:1997bh}. The matrix models resulting from supersymmetric localization in the gauge theory deformations will be evaluated using numerical saddle point techniques, for a range of finite $N$ for the $T_N$ theories and finite $N$, $M$ for the $\#_{N,M}$ theories. These saddle point techniques are expected to become exact at large $N$ and large $(N,M)$ for the $T_N$ and $\#_{N,M}$ theories, respectively.
From the numerical saddle point results we can thus reliably extract the leading-order terms in the partition functions and conformal central charges at large $N$ and large $(N,M)$, and compare to the results obtained from holographic analyses using the solutions that are proposed to correspond to these theories.

The outline of the remaining part is as follows. In Section~\ref{sec:2} we discuss gauge theory deformations of the five-dimensional $T_N$ and $\#_{N,M}$ superconformal field theories. In Section~\ref{sec:matrixmodels} we set up the corresponding matrix models, which compute the squashed five-sphere partition functions from supersymmetric localization, and compute the five-sphere partition functions as well as the conformal central charges $C_T$ using numerical saddle point methods. In Section~\ref{sec:3} we review the supergravity solutions associated with the five-brane webs of the superconformal field theories and discuss the holographic results for the partition functions. A comparison of the results is presented in sec.~\ref{sec:comp}, and we end with a discussion in sec.~\ref{sec:4}. Technical details and additional numerical results are contained in two appendices.

\section{The five-dimensional \texorpdfstring{$T_N$}{T-N} and \texorpdfstring{$\#_{N,M}$}{\#-N-M} theories}
\label{sec:2}

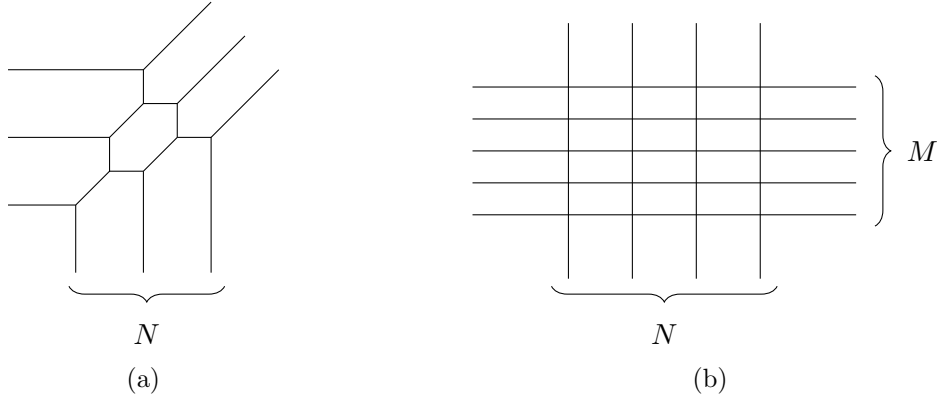
\begin{figure}
\centering
\subfloat[][]{\label{fig:TN-web}
 \begin{tikzpicture}
   \begin{scope}[scale=0.9]
   \draw (-3/2,-1/2) -- (-1/2,-1/2) -- (0,0) -- (1/2,0) -- (1,1/2) -- (1,1) -- (1/2,1) -- (0,1/2) -- (0,0);
   \draw (-1/2,-1/2) -- (-1/2,-3/2);
   \draw (0,1/2) -- (-3/2,1/2);
   \draw (1/2,0) -- (1/2,-3/2);
   \draw (1,1) -- (2,2);
   
   \draw (1,1/2) -- (3/2,1/2) -- (3/2,-3/2);
   \draw (3/2,1/2) -- (5/2,3/2);
   
   \draw (1/2,1) -- (1/2,3/2) -- (3/2,5/2);
   \draw (1/2,3/2) -- (-3/2,3/2);
   
   \draw [decorate,decoration={brace,amplitude=6pt},xshift=0pt,yshift=0pt]
(1.7,-1.7) -- (-0.6,-1.7)node [black,midway,yshift=-18pt] {\small $N$};
\end{scope}
 \end{tikzpicture}
}
\hskip 0.9in
\subfloat[][]{\label{fig:D5NS5-web}
\begin{tikzpicture}
\begin{scope}[scale=0.85]
 \foreach \i in {-2,...,2}{
  \draw (-3,\i/2) -- (3,\i/2);
 }
 \foreach \i in {-3/2,-1/2,1/2,3/2}{
  \draw (\i,-2) -- (\i,2);
 }
\end{scope}
\draw [decorate,decoration={brace,amplitude=6pt},xshift=0pt,yshift=0pt]
(1.5,-1.8) -- (-1.5,-1.8)node [black,midway,yshift=-18pt] {\small $N$};

\draw [decorate,decoration={brace,amplitude=6pt},xshift=0pt,yshift=0pt]
(2.8,1.0) -- (2.8,-1.0)node [black,midway,xshift=18pt] {\small $M$};
\end{tikzpicture}
}
 \caption{Five-brane webs for the $T_N$ theory with $N=3$ on the left hand side and for the $\#_{N,M}$ theory with $N=5$, $M=4$ on the right hand side. The brane webs show relevant deformations of the superconformal field theories.\label{fig:webs}}
\end{figure}

In this section we introduce the two five-dimensional superconformal field theories whose partition functions and central charges we will compute and compare with the corresponding supergravity results. We discuss their realization as $(p,q)$ five-brane webs in Type IIB string theory, which allows for the computation of their partition function from supergravity, as well as their deformations to infrared gauge theories, which permit a field theory computation of the partition functions using supersymmetric localization.

The first example are the five-dimensional $T_N$ theories~\cite{Benini:2009gi,Bao:2013pwa}. They are realized on an intersection of $N$ D5-branes, $N$ NS5-branes and $N$ $(1,1)$ five-branes. A brane web corresponding to a deformation away from the superconformal fixed point to the gauge theory phase is shown in Figure~\ref{fig:TN-web}. The ultraviolet superconformal field theory is retained in the limit where all external five-branes intersect at a single point. Upon compactification on $\rm S^1$, these theories reduce to the four-dimensional $T_N$ theories~\cite{Gaiotto:2009we} (see~\cite{Tachikawa:2015bga} for a nice review). A gauge theory description for the five-dimensional theories is given by the linear quiver~\cite{Bergman:2014kza,Hayashi:2014hfa}
\ie\label{eq:TN-quiver}
\raisebox{-0.7cm}{\includegraphics[width=.85\textwidth]{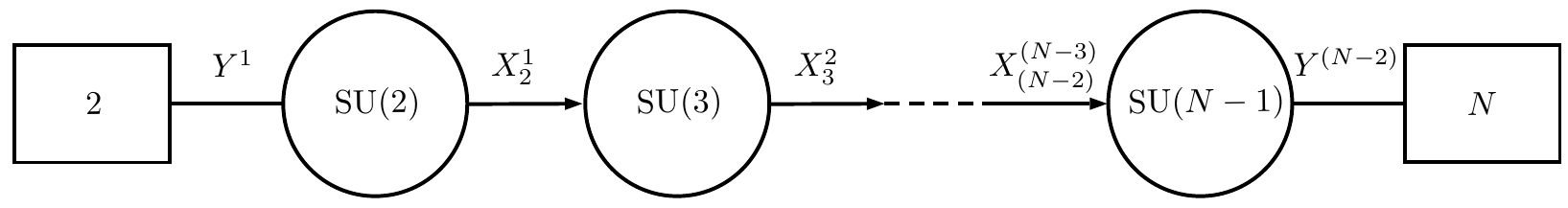}}
\fe
The circles denote gauge group nodes with the gauge group as label, while the squares denote hypermultiplets in the fundamental representation of the gauge group node they are attached to. The $X^{i}_{i+1}$ denote bifundamental hypermultiplets between adjacent gauge group nodes, and by $Y^1$, $Y^{(N-2)}$ we denote the two, $N$ fundamental hypermultiplets for the first, last gauge node, respectively. Each of the bifundamental hypermultplets gives rise to a ${\rm U}(1)_{\rm b}$ symmetry, and each gauge node gives rise to a ${\rm U}(1)_{\rm I}$ instanton symmetry. The resulting global symmetry group of the gauge theory is given by
\ie
G_{F,{\rm IR}}^{T_N} \ = \ {\rm SO}(4)\times {\rm U}(1)_{\rm I}^{N-2} \times {\rm U}(1)_{\rm b}^{N-3} \times {\rm U}(N)\,,
\fe
where the first and last factor are the flavor symmetries associated with the fundamental fields. The Chern-Simons levels are zero for all nodes~\cite{Bergman:2014kza}.
As is evident from the brane web, the ultraviolet superconformal field theory has (enhanced) global symmetry
\ie
G_{F,{\rm UV}}^{T_N} \ = \ {\rm SU}(N)\times {\rm SU}(N)\times{\rm SU}(N)\,.
\fe

The second theory in consideration in this paper is realized by an intersection of $N$ D5-branes and $M$ NS5-branes. This theory has originally been discussed in~\cite{Aharony:1997bh}, and a corresponding brane web is shown in Figure~\ref{fig:D5NS5-web}. Inspired by the shape of the web we refer to this theory as the $\#_{N,M}$ theory. In this case, a relevant deformation flowing to a gauge theory in the infrared yields the linear quiver
\ie\label{eq:HMN-quiver}
 \raisebox{-1.1cm}{\includegraphics[width=.85\textwidth]{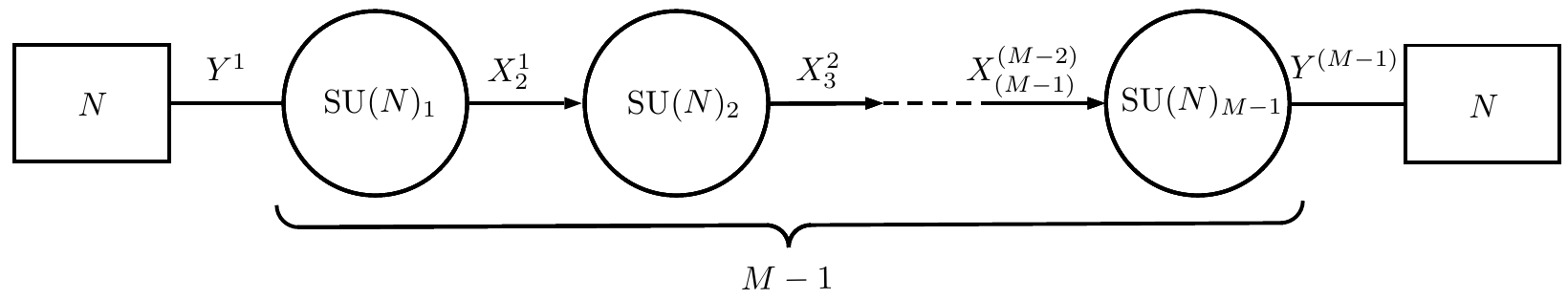}}
\fe
where the $SU(N)$ node appears $(M-1)$ times. The Chern-Simons levels are zero for all nodes. We again use $X^{i}_{i+1}$ to denote bifundamental hypermultiplets between adjacent gauge group nodes, and $Y^{1}$, $Y^{M-1}$ to denote $N$ fundamental hypermultiplets for the first, last node, respectively. The infrared gauge theory exhibits the ${\rm U}(N)$ flavor symmetries arising from the $N$ fundamentals at each end, the ${\rm U}(1)_{\rm b}$ symmetries associated with the bifundamental hypermultiplets, as well as the instanton ${\rm U}(1)_{\rm I}$ symmetries,
\ie
G_{F,{\rm IR}}^{\#_{N,M}} \ = \ {\rm U}(N) \times {\rm U}(1)_{\rm b}^{M-2} \times {\rm U}(1)_{\rm I}^{M-1} \times {\rm U}(N)\,.
\fe
This infrared symmetry is enhanced at the ultraviolet superconformal fixed point to\footnote{For small $M$, $N$ the symmetry can be further enhanced; \emph{e.g.}, for $N=2$ it contains an ${\rm SU}(2M)$-factor~\cite{Aharony:1997bh}.}
\ie
G_{F,{\rm UV}}^{\#_{N,M}} \ = \ {\rm SU}(N)\times {\rm SU}(N) \times {\rm SU}(M)\times {\rm SU}(M)\times {\rm U}(1) \,.
\fe
The S-dual quiver is obtained by rotating the brane web by 90 degrees, and is given by
\ie\label{eq:HMN-dual-quiver}
 \raisebox{-1.1cm}{\includegraphics[width=.85\textwidth]{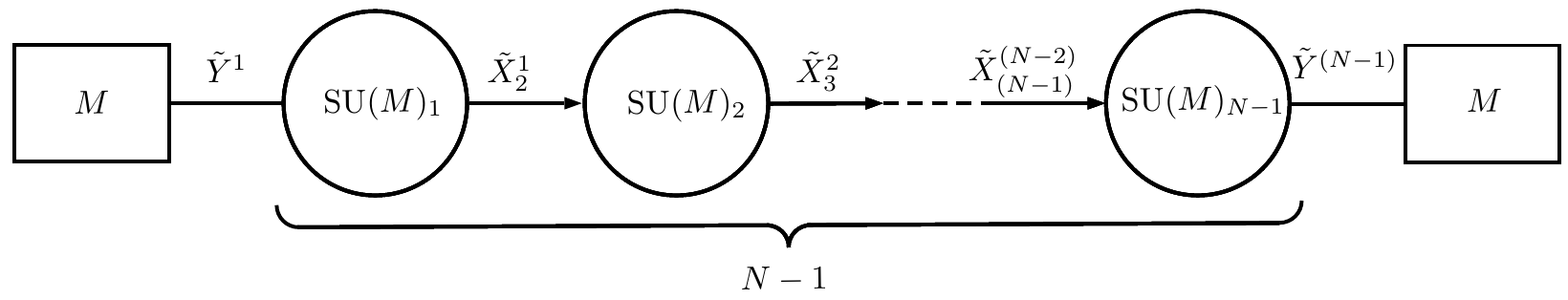}}
\fe
where the $SU(M)$ node now appears $(N-1)$ times.

\section{Partition functions and central charges from localization}\label{sec:matrixmodels}

In this section we spell out the matrix models for squashed five-sphere partition functions for the gauge theory deformations of the five-dimensional $T_N$ and $\#_{N,M}$ theories introduced in the previous section. The infrared description in terms of a conventional Lagrangian gauge theory allows for localization computations. Generically, supersymmetric localization computations using a supercharge $\cQ$ are sensitive to all operators in the cohomology of $\cQ$. One may therefore expect computations for the ultraviolet conformal field theory to receive contributions from higher-derivative terms in the infrared Lagrangian description. 
However, it was  conjectured in~\cite{Kim:2012ava,Jafferis:2012iv,Kim:2012qf} that 
all such contributions are $\cQ$-exact, and thus will not affect the localization computation. This argument suggests that the result for the superconformal field theory partition function can be obtained by simply taking the (bare) five-dimensional Yang-Mills coupling to infinity (\emph{i.e.} $g_{\rm YM} \to \infty$) in the infrared gauge theory localization.

Generically, (squashed) five-sphere partition functions receive instanton contributions, arising from localization in nontrivial instanton backgrounds. For the superconformal field theory at $g_{\rm YM} \to \infty$ (or equivalently at vanishing instanton mass $m_{\rm I} \to 0$), terms of all instanton numbers contribute equally, and na\"ively a perturbative expansion is not possible. However, it was argued in~\cite{Jafferis:2012iv} that at large $N$ a simplification occurs, in that the instanton contributions are in fact exponentially suppressed. At the heart of the argument lies the fact that the instanton mass $m_{\rm I} \propto 1/g_{\rm YM}^{2}$ is BPS and thus  determined by the central charge $Z$ of the superconformal algebra. Hence, the instanton expansion is actually controlled by $Z$, which in turn is related to the effective Yang-Mills coupling $g_{\rm eff}$, given by
\ie
\frac{1}{ g ^{2}_{\rm eff}} \ \sim \ \frac{1}{g_{\rm YM}^{2}} + \left| \lambda \right| \,,
\fe
where we collectively denote the Coulomb branch parameters by $\lambda$~\cite{Seiberg:1996bd,Morrison:1996xf,Intriligator:1997pq}. As we explicitly (numerically) observe for the theories in consideration, the Coulomb branch parameters $\lambda$ scale as\footnote{The gauge theories for both superconformal field theories are infinite quivers in the strict large $N$, $(M,N)$ limits, and the scaling exponents are not universal. However, they are always positive, $\alpha,\beta >0$.}
\ie
\lambda^{(T_N)} \ \sim \ N^{\alpha} \,, \quad \alpha >0\,,
\fe
for $T_N$ theories, and as
\ie
\lambda^{(\#_{N,M})} \ \sim \ (MN)^{\beta} \,, \quad \beta >0 \,,
\fe
for $\#_{N,M}$ theories. Thus, we expect that in the respective limits, the inverse effective Yang-Mills coupling diverges, and as a consequence the contributions with instanton numbers $n>0$ are exponentially suppressed in the partition function. 

We further notice that in the case of Seiberg theories it was (experimentally) observed in~\cite{Chang:2017cdx} that even for small $N$, instanton contributions to the five-sphere free energy are small compared to the zero-instanton localization. Since Seiberg theories with $N_{\rm f}=4$ and $N_{\rm f}=3$ are equivalent to $T_2$ and $\#_{2,2}$, respectively, this result suggests that the zero-instanton contribution approximates the full five-sphere partition function well even for $T_N$ and $\#_{N,M}$ theories at small $N$ and small $(M,N)$, respectively.

\subsection{Five-dimensional \texorpdfstring{$T_N$}{T-N} theories}

We now spell out the perturbative squashed five-sphere partition function for the gauge theory deformations of the $T_N$ superconformal field theories introduced in Section~\ref{sec:2}. We refer to Appendix~\ref{App:PartFunc} for details on the general form of the perturbative (zero-instanton) part of the squashed five-sphere partition function for arbitrary theories. For notational convenience we define
\bea\label{eqn:ZTN}
\cZ^{T_N}_{\mathrm{pert}} \left[ \omega_1, \omega_2, \omega_3 \right]\ = \ \int_{-\infty}^{\infty} \left[ \prod_{j=2}^{N-1} \prod_{i=1}^{j-1} \diff \lambda_{i}^{(j)}  \right]\exp \left( - \cF^{T_N}_{[\omega_1,\omega_2,\omega_3]} \big(  \lambda_{i}^{(j)} \big) \right) \,.
\eea
With the simplifying notation 
\begin{align}\label{eqn:S3Onot}
S_{3} (z) \ &\equiv \ S_{3}\left( z \mid \omega_1, \omega_2, \omega_3 \right) \,, & 
\omega_{\rm tot} \ &\equiv \ \omega_1+\omega_2+\omega_3\,,
\end{align}
the exponent $\cF^{T_N}_{[\omega_1,\omega_2,\omega_3]}\big(  \lambda_{i}^{(j)} \big)$ is explicitly given by  
\ie\label{eqn:cFTN}
\cF^{T_N}_{[\omega_1,\omega_2,\omega_3]} \big(  \lambda_{i}^{(j)} \big)  \ = \ & 
- \frac{(N-2)(N-1)}{2} \log \frac{S_{3}^{\prime} (0)}{2\pi}
+ \sum_{j=2}^{N-1} \log j! \\
&
-\sum_{j=2}^{N-1} \sum_{\substack{\ell,m =1 \\ \ell \neq m}}^{j} \frac{1}{2}\log S_3 \left( \pm \ii \left[ \lambda_{\ell}^{(j)} -\lambda_{m}^{(j)} \right]  \right) \\
&
+\sum_{j=2}^{N-1}\left[\sum_{\ell=1}^{j}\sum_{m=1}^{j+1}   \log S_3 \left( \ii \left[ \lambda_{\ell}^{(j)} - \lambda_{m}^{(j+1)} \right] + \frac{\omega_{\rm tot}}{2} \right) \right] \\
&
+ 2 \sum_{i=1}^{2} \log S_3 \left( \ii \lambda_{i}^{(2)} +\frac{\omega_{\rm tot}}{2} \right) 
+ N \sum_{i=1}^{N-1}  \log S_3 \left( \ii \lambda_{i}^{(N-1)}+\frac{\omega_{\rm tot}}{2} \right) \,,
\fe
where we further impose the following constraint on the Coulomb branch parameters $\lambda_i^{(j)}$
\bea\label{eqn:constTN}
\sum_{i=1}^{j} \lambda_{i}^{(j)} \ = \ 0 \,, \quad \forall j \in \{2, \ldots, N-1\} \,.
\eea
Notice that, to describe the superconformal field theory partition function, we take the five-dimensional Yang-Mills coupling to infinity, $1/g^{2}_{YM}  = 0$. Furthermore, possible Chern-Simons couplings are turned off for the theories in consideration.

\subsection{Five-dimensional \texorpdfstring{$\#_{N,M}$}{D5NS5} theories}

Now let us turn to the $\#_{N,M}$ theories, for which a gauge theory deformation is given in equation~(\ref{eq:HMN-quiver}). Again, we start by rewriting the partition function of the quiver theory as
\bea\label{eqn:ZMN}
\cZ^{\#_{N,M}}_{{\rm pert}} \left[ \omega_1,\omega_2,\omega_3 \right] \ = \ \int_{-\infty}^{\infty} \left[ \prod_{j=1}^{M-1} \prod_{i=1}^{N-1} \diff \lambda_{i}^{(j)}  \right]
\exp \left( - \cF^{\#_{N,M}}_{[\omega_1,\omega_2,\omega_3]}\big(  \lambda_{i}^{(j)} \big) \right) \,,
\eea
where, using the notation in~\eqref{eqn:S3Onot}, the exponent $\cF^{\#_{N,M}}_{[\omega_1,\omega_2,\omega_3]}\big(  \lambda_{i}^{(j)} \big)$ is explicitly given by
\ie\label{eqn:cFMN}
\cF^{\#_{N,M}}_{[\omega_1,\omega_2,\omega_3]}\big(  \lambda_{i}^{(j)} \big) \ = \ 
& 
- (N-1)(M-1) \log \frac{S_{3}^{\prime} (0)}{2\pi} \\
&
- \sum_{j=1}^{M-1} \sum_{\substack{\ell,m =1 \\ \ell \neq m}}^{N} \frac{1}{2}\log S_{3}\left( \pm \ii \left[ \lambda_{\ell}^{(j)} -\lambda_{m}^{(j)} \right]  \right) \\
&
+\sum_{j=1}^{M-2} \sum_{\ell,m=1}^{N} \log S_{3}\left( \ii \left[ \lambda_{\ell}^{(j)} - \lambda_{m}^{(j+1)} \right]+\frac{\omega_{\rm tot}}{2} \right) \\
&
+N \sum_{i=1}^{N} \left[ \log S_{3}\left(\ii \lambda_{i}^{(1)} + \frac{\omega_{\rm tot}}{2} \right) +\log S_{3}\left( \ii \lambda_{i}^{(M-1)} + \frac{\omega_{\rm tot}}{2} \right)  \right] \,.
\fe
As before, we have to further impose the additional constraint
\bea\label{eqn:constHT}
\sum_{i=1}^{N} \lambda_{i}^{(j)} \ = \ 0 \,, \quad \forall j \in \left\{1, \ldots, M-1\right\} \,,
\eea
for the Coulomb branch parameters. Furthermore, the inverse five-dimensional Yang-Mills coupling $1/g_{\rm YM}^{2}$ vanishes and possible Chern-Simons terms are turned off.

\subsection{Strategy for numerical evaluation}\label{sec:largeN}

We now turn towards explicitly evaluating the perturbative parts of the (squashed) five-sphere partition function for the two classes of theories in consideration. To do so, we have to explicitly evaluate the Coulomb branch integrals in~\eqref{eqn:ZTN} and~\eqref{eqn:ZMN} for the matrix models of equations~\eqref{eqn:cFTN} and~\eqref{eqn:cFMN}. We do so by employing a numerical saddle point method. On general grounds, a saddle point evaluation will be exact only in the strict parametrically large $N$, $(M,N)$ limits, and these limits will be our main interest here. However, we will also present consistency checks suggesting that these results can be taken seriously even for relatively small parameters. This was already observed in~\cite{Chang:2017mxc} in the case of Seiberg theories.

We follow the numerical method of~\cite{Herzog:2010hf} to evaluate the Coulomb branch integrals using a saddle point approximation. The saddle point equations are explicitly given by
\begin{align}\label{eqn:saddlepointeqnTN}
\frac{\partial \cF_{[\omega_1,\omega_2,\omega_3]}^{T_N}}{\partial \lambda_i^{(j)}} \ &= \  0 \,, & \forall i &= 1, \ldots, j-1 \,, \ j=1, \ldots N-1 \,,\\
\label{eqn:saddlepointeqnHT}
\frac{\partial \cF_{[\omega_1,\omega_2,\omega_3]}^{\#_{N,M}}}{\partial \lambda_i^{(j)}} \ &= \   0 \,, & \forall j &= 1, \ldots, M-2 \,,\ i=2, \ldots N-1 \,,
\end{align}
where we have imposed the constraints~\eqref{eqn:constTN} and~\eqref{eqn:constHT}. Though highly non-trivial, these equations can be solved numerically. To do so, it is convenient to rephrase the problem in terms of a system of $\frac{(N-2)(N-1)}{2}$ particles in the case of~\eqref{eqn:saddlepointeqnTN} and $(N-1)(M-1)$ particles in the case of~\eqref{eqn:saddlepointeqnHT} with time-dependent coordinates $\lambda_i^{(j)}( t )\in \mathbb{R}$ moving in a potential given by the explicit expressions~\eqref{eqn:constTN} and~\eqref{eqn:constHT}, respectively. The corresponding equations of motion of that system of particles then read
\begin{align}
\frac{\partial \cF_{[\omega_1,\omega_2,\omega_3]}^{T_N}}{\partial \lambda_i^{(j)}(t)} \ &= \  \xi_{T_N} \frac{\diff \lambda_i^{(j)}(t)}{\diff t} \,, & \forall i &= 1, \ldots, j-1 \,, \ j=1, \ldots N-1 \,,\\
\frac{\partial \cF_{[\omega_1,\omega_2,\omega_3]}^{\#_{N,M}}}{\partial \lambda_i^{(j)}(t)} \ &= \  \xi_{\#_{N,M}} \frac{\diff \lambda_i^{(j)}(t)}{\diff t} \,, & \forall j &= 1, \ldots, M-2 \,,\ i=2, \ldots N-1 \,,
\end{align}
where we have included signs $\xi_{T_N} \in \{\pm1\}$ and $\xi_{\#_{N,M}}\in \{\pm1\}$ to ensure that the saddle points are attractive; in our case $\xi_{T_N}=\xi_{\#_{N,M}} = -1$. Then, at large times $t \to \infty$, the equilibrium configurations of these particles describe solutions to the saddle point equations~\eqref{eqn:saddlepointeqnTN} and~\eqref{eqn:saddlepointeqnHT}.

With the solutions to the saddle point equations we can then evaluate the Coulomb branch integrals for the five-sphere partition functions. As derived in~\cite{Chang:2017cdx}, and reviewed in Appendix~\ref{App:CT}, the conformal central charge $C_T$ is related to squashing deformations of the five-sphere partition function, explicitly detailed in equation~\eqref{eqn:CTvsF}. Thus, by evaluating the leading-order corrections to the round sphere partition functions in terms of squashing parameters, we can explicitly compute $C_T$ using the saddle point approximation. 

The saddle point approximation is expected to become exact in the limit where the parameters $N$ and $(M,N)$ for the $T_N$ and $\#_{N,M}$ theories, respectively, are large. At parametrically small $N$, $(M,N)$, one may expect large relative corrections in the explicitly evaluated perturbative part of the full five-sphere partition function. Furthermore, one may expect instanton contributions. However, as observed in~\cite{Chang:2017mxc}, for Seiberg theories, the explicitly evaluated perturbative five-sphere partition function (and some central charges) are actually very accurately reproduced by the saddle point method even at small $N$, with relative errors of less than $1.6 \%$. The $T_2$, $\#_{2,2}$ theories are equivalent to rank-one Seiberg theories with $N_{\rm f} = 3$, $N_{\rm f} = 4$, respectively, and the respective errors are below $0.5\%$. This suggests that the saddle point approximation accurately captures the actual results also for the $T_N$ and $\#_{N,M}$ theories. We will in the following discuss the explicit numerical results and provide additional support to this effect by comparing the five-sphere free energy of S-dual theories $\#_{N,M}$ and $\#_{M,N}$.

\subsection{Results}\label{sec:num-results}

\begin{table}
\centering
\setlength{\tabcolsep}{7.5pt}
\renewcommand{\arraystretch}{1.2}
\begin{tabular}{c|c||c|c||c|c||c|c||c|c}
\toprule
 $N$  & $ -F_{\rm S^5}^{T_{N}}$ &
 $N$  & $ -F_{\rm S^5}^{T_{N}}$ &
 $N$  & $ -F_{\rm S^5}^{T_{N}}$ &
 $N$  & $ -F_{\rm S^5}^{T_{N}}$ &
  $N$  & $ -F_{\rm S^5}^{T_{N}}$ 
 \\
\hline
$ 3 $ & $ 13.8861 $ & $ 13 $ & $ 10796.5 $ & $ 23 $ & $ 110596. $ & $ 33 $ & $ 475466. $ & $ 43 $ & $ 1.38015\cdot 10^6 $ \\
$ 4 $ & $ 63.3184 $ & $ 14 $ & $ 14639.6 $ & $ 24 $ & $ 131395. $ & $ 34 $ & $ 536254. $ & $ 44 $ & $ 1.51383\cdot 10^6 $ \\
$ 5 $ & $ 181.527 $ & $ 15 $ & $ 19423.0 $ & $ 25 $ & $ 154993. $ & $ 35 $ & $ 602689. $ & $ 45 $ & $ 1.65700\cdot 10^6 $ \\
$ 6 $ & $ 411.093 $ & $ 16 $ & $ 25288.2 $ & $ 26 $ & $ 181632. $ & $ 36 $ & $ 675108. $ & $ 46 $ & $ 1.81038\cdot 10^6 $ \\
$ 7 $ & $ 804.544 $ & $ 17 $ & $ 32386.7 $ & $ 27 $ & $ 211560. $ & $ 37 $ & $ 753861. $ & $ 47 $ & $ 1.97351\cdot 10^6 $ \\
$ 8 $ & $ 1424.33 $ & $ 18 $ & $ 40879.6 $ & $ 28 $ & $ 245038. $ & $ 38 $ & $ 839306. $ & $ 48 $ & $ 2.14777\cdot 10^6 $ \\
$ 9 $ & $ 2342.81 $ & $ 19 $ & $ 50938.1 $ & $ 29 $ & $ 282336. $ & $ 39 $ & $ 931811. $ & $ 49 $ & $ 2.33332\cdot 10^6 $ \\
$ 10 $ & $ 3642.20 $ & $ 20 $ & $ 62743.2 $ & $ 30 $ & $ 323733. $ & $ 40 $ & $ 1.03176\cdot 10^6 $ & $ 50 $ & $ 2.53117\cdot 10^6 $ \\
$ 11 $ & $ 5414.63 $ & $ 21 $ & $ 76485.8 $ & $ 31 $ & $ 369519. $ & $ 41 $ & $ 1.13953\cdot 10^6 $ & $ 51 $ & $ 2.74019 \cdot 10^6 $ \\
$ 12 $ & $ 7762.10 $ & $ 22 $ & $ 92366.6 $ & $ 32 $ & $ 419994. $ & $ 42 $ & $ 1.25552\cdot 10^6 $ & $ 52 $ & $ 2.96251\cdot 10^6 $ \\
\bottomrule
\end{tabular}
\caption{Explicit numerical values for the saddle point evaluation of the five-sphere free energy for five-dimensional $T_{N}$ theories for $3 \leq N \leq 52$.}\label{tbl:TNF-main}
\end{table}

\begin{figure}
\centering
 \subfloat{
   \includegraphics[width=.48\textwidth]{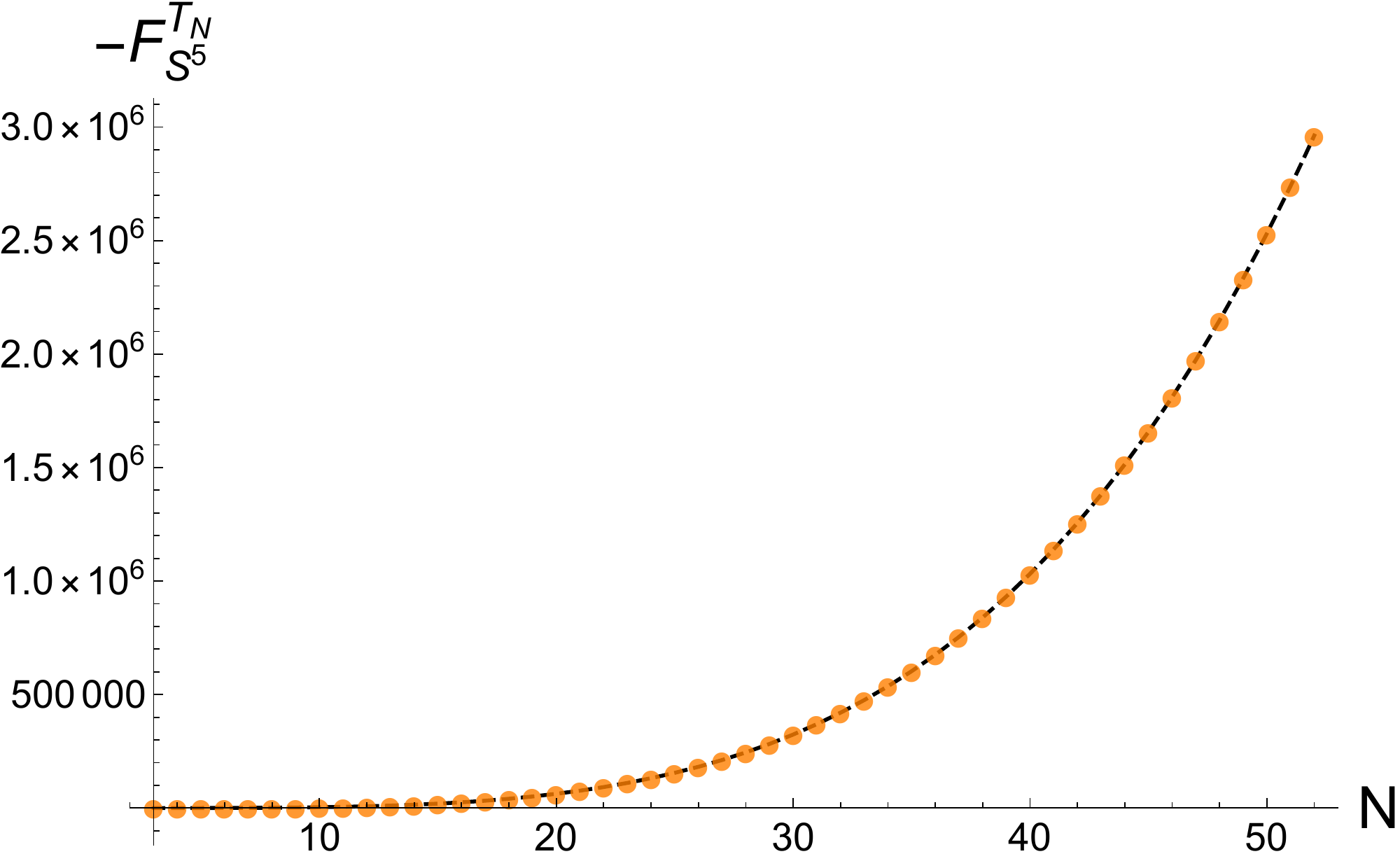}
 }\hskip -0.0in
 \subfloat{
 \includegraphics[width=.48\textwidth]{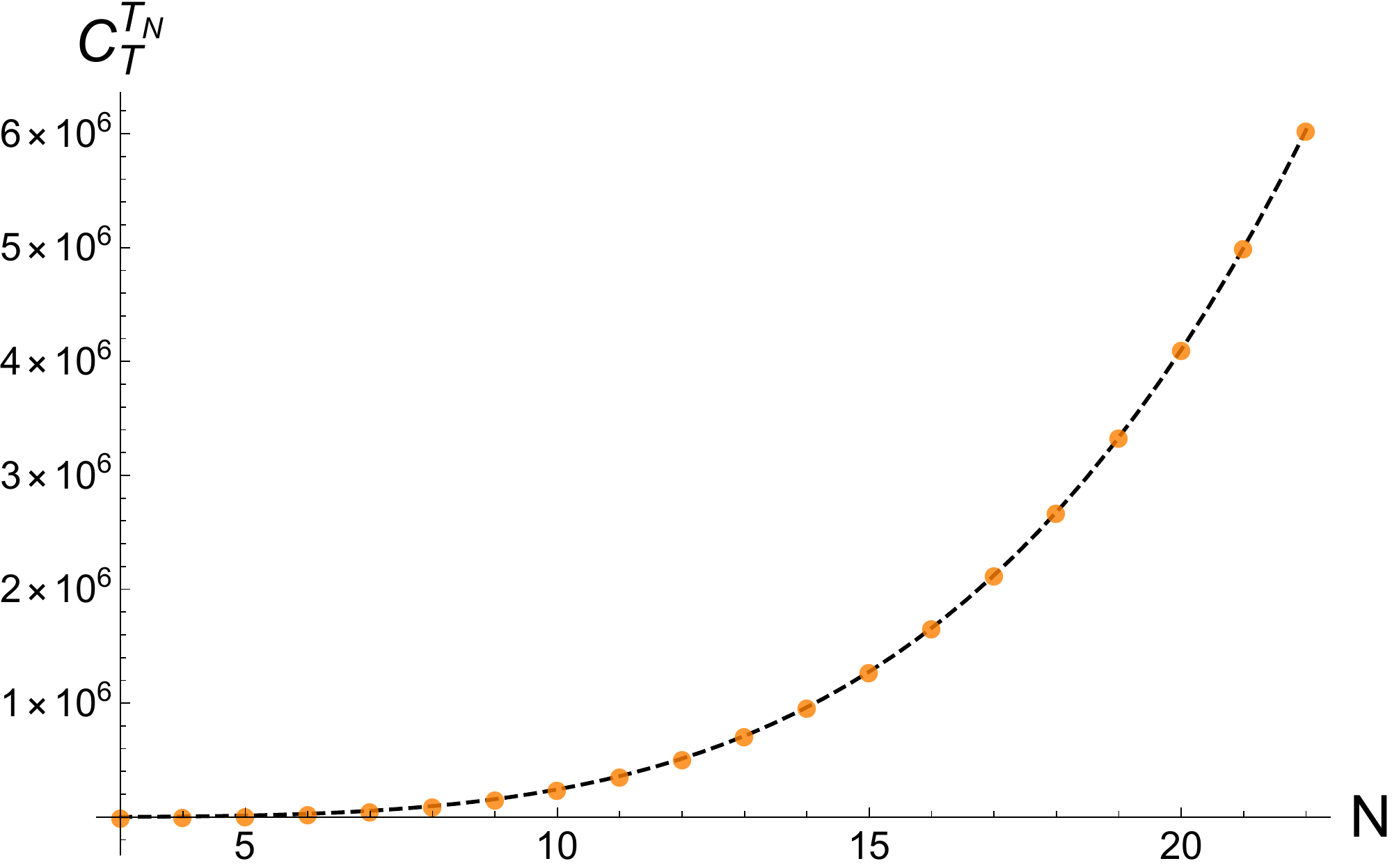}
 }
 \caption{The left hand side shows an overlay of the numerical results for the five-sphere partition functions $-F_{\rm S^{5}}^{T_N}$ of the $T_N$ theories with the least-squares fit to a degree four polynomial. The right hand side depicts the same for the conformal central charge $C_T^{T_N}$.\label{fig:TN-fits}}
\end{figure}

In this section we present the results and analyses of the numerical evaluation of the five-sphere partition functions and conformal central charges for the $T_N$ and $\#_{N,M}$ theories. The explicit results for five-sphere free energy, $F^{T_N}=-\log\mathcal Z^{T_N}$, for the $T_N$ theories, for $3\leq N\leq 52$, are shown in Table~\ref{tbl:TNF-main}. This data shows striking agreement with a quartic scaling ansatz; a least-squares fit to a degree four polynomial in $N$,
\ie\label{eq:F-TN-fit}
 -F_{\rm fit}^{T_N}& \ = \ \sum_{i=0}^4 a_i N^i\,,
\fe
yields the following result
\begin{align}
 a_4&  \ = \ 0.411184\,, & a_3& \ = \ -0.257119\,,
 &a_2 \ = \ -3.06860\,, \nonumber\\
 a_1& \ = \  13.5718\,, & a_0& \ = \ -35.8536\,.
\end{align}
The relative error, 
\ie
R \ \equiv \ \left| \frac{F_{\rm S^5}^{T_{N}}-F_{\rm fit}^{T_N}}{F_{\rm S^5}^{T_{N}}}\right|~,
\fe
is $\mathcal O(10^{-3})$ or below for all $N>10$. For small $N$ one may expect large deviations from a simple scaling ansatz, but the deviations are small even for small values of $N$: for $5\leq N\leq 10$, the relative error is still less than $1\%$, and only for $N=3$ and $N=4$ do we see substantial relative errors. The results are illustrated in Figure~\ref{fig:TN-fits}. We emphasize that for the comparison to the supergravity result only the scaling of the leading large $N$ term and its coefficient are relevant. It is a perhaps curious observation how well the polynomial ansatz including the subleading terms in~(\ref{eq:F-TN-fit}) reproduces our numerical results. On general grounds, there may well be additional terms, and the coefficients of the subleading terms may be subject to sizable corrections. These effects are, however, irrelevant for the comparison to the supergravity result.

\begin{table}
\centering
\setlength{\tabcolsep}{9pt}
\renewcommand{\arraystretch}{1.2}
\begin{tabular}{c|c||c|c||c|c||c|c}
\toprule
 $N$  & $ C_T^{T_{N}}$ &
 $N$  & $ C_T^{T_{N}}$ &
 $N$  & $ C_T^{T_{N}}$ &
 $N$  & $ C_T^{T_{N}}$ 
 \\
\hline
$ 3 $ & $ 1109.23 $ & $ 8 $ & $ 95768.6 $ & $ 13 $ & $ 711839. $ & $ 18 $ & $ 2.67706\cdot 10^6 $ \\
$ 4 $ & $ 4620.53 $ & $ 9 $ & $ 156542. $ & $ 14 $ & $ 963424. $ & $ 19 $ & $ 3.33300\cdot 10^6 $ \\
$ 5 $ & $ 12738.5 $ & $ 10 $ & $ 242234. $ & $ 15 $ & $ 1.27624\cdot 10^6 $ & $ 20 $ & $ 4.10247\cdot 10^6 $ \\
$ 6 $ & $ 28248.0 $ & $ 11 $ & $ 358827. $ & $ 16 $ & $ 1.65946\cdot 10^6 $ & $ 21 $ & $4.99786\cdot 10^6 $\\
$ 7 $ & $ 54570.9 $ & $ 12 $ & $ 512941. $ & $ 17 $ & $ 2.12291\cdot 10^6 $ & $ 22 $ & $ 6.03219\cdot 10^6$ \\
\bottomrule
\end{tabular}
\caption{Explicit numerical values for $C_T$ of five-dimensional $T_{N}$ theories computed using the numerical saddle point evaluation.}\label{tbl:TNCT-main}
\end{table}

Similar statements apply for the conformal central charge $C_T$. The results obtained from the numerical evaluation of the deformed five-sphere partition function are shown in Table~\ref{tbl:TNCT-main}. Since the evaluation is more time-consuming, we restricted to $N\leq 22$. We perform a similar least-squares fit of the data to a degree-four polynomial,
\begin{align}
 C_{T,{\rm fit}}^{T_N}& \ = \ \sum_{i=0}^4 b_i N^i\,,
\end{align}
which yields
\begin{align}
 b_4& \ = \ 26.6647\,, & b_3& \ = \ -16.2736\,, 
 &b_2 \ = \ -87.0646\,, \nonumber\\
 b_1& \ = \  56.3415\,, & b_0& \ = \ 2.97664\,.
\end{align}
Again, this provides a remarkably good approximation, with relative errors less than $\cO(10^{-3})$. The combined results are illustrated in Figure~\ref{fig:TN-fits}.

For the $\#_{N,M}$ theories, with two independent parameters, we tabulate the explicit numerical results for $F^{\#_{N,M}}=-\log\mathcal Z^{\#_{N,M}}$, obtained for $2\leq N,M\leq 30$, in Appendix~\ref{App:Numerics}. We focus here on a discussion of the physical features. As a first step, we quantify the discrepancy between the results for the quiver deformation~(\ref{eq:HMN-quiver}) and for the S-dual quiver deformation~(\ref{eq:HMN-dual-quiver}). This will provide a nontrivial consistency check on our computations. As discussed in the previous section, instanton contributions are expected to be suppressed in the large $N,M$ limit. The comparison between the quiver and S-dual quiver will therefore serve as a quantitative indicator for the impact of instantons, and for the ``error" of neglecting them. We define
\begin{align}\label{eqn:etadef}
 \eta_{N,M}& \ = \ \left| \frac{F_{\rm S^5}^{\#_{N,M}}-F_{\rm S^5}^{\#_{M,N}}}{F_{\rm S^5}^{\#_{N,M}}}\right|\,.
\end{align}
The discrepancies are generally small around the diagonal, where $N\approx M$, and grow with $\left|N-M\right|$. The largest discrepancies for $N<M$ and $N>M$ occur, not surprisingly, for $(N,M)=(2,30)$ and $(N,M)=(30,2)$, respectively, and are of order $10\%$. The crucial point for the analysis to be carried out here is that, as expected, these discrepancies indeed decrease as $N,M$ increase. For example, restricting to 
\begin{align}\label{eq:hash-restriction}
 16 \ \leq \ M, N  \ \leq \ 30
\end{align}
reduces the maximal discrepancy to less than $1\%$ for maximal $|N-M|$. For the larger $(N,M)$ values the discrepancy is an order of magnitude smaller. We refer to Figure~\ref{Fig:Sduality} for a plot depicting the convergence properties for a sample of $\eta_{M,N}$'s.

\begin{figure}
 \centering
 \subfloat{
  \includegraphics[width=0.49\linewidth]{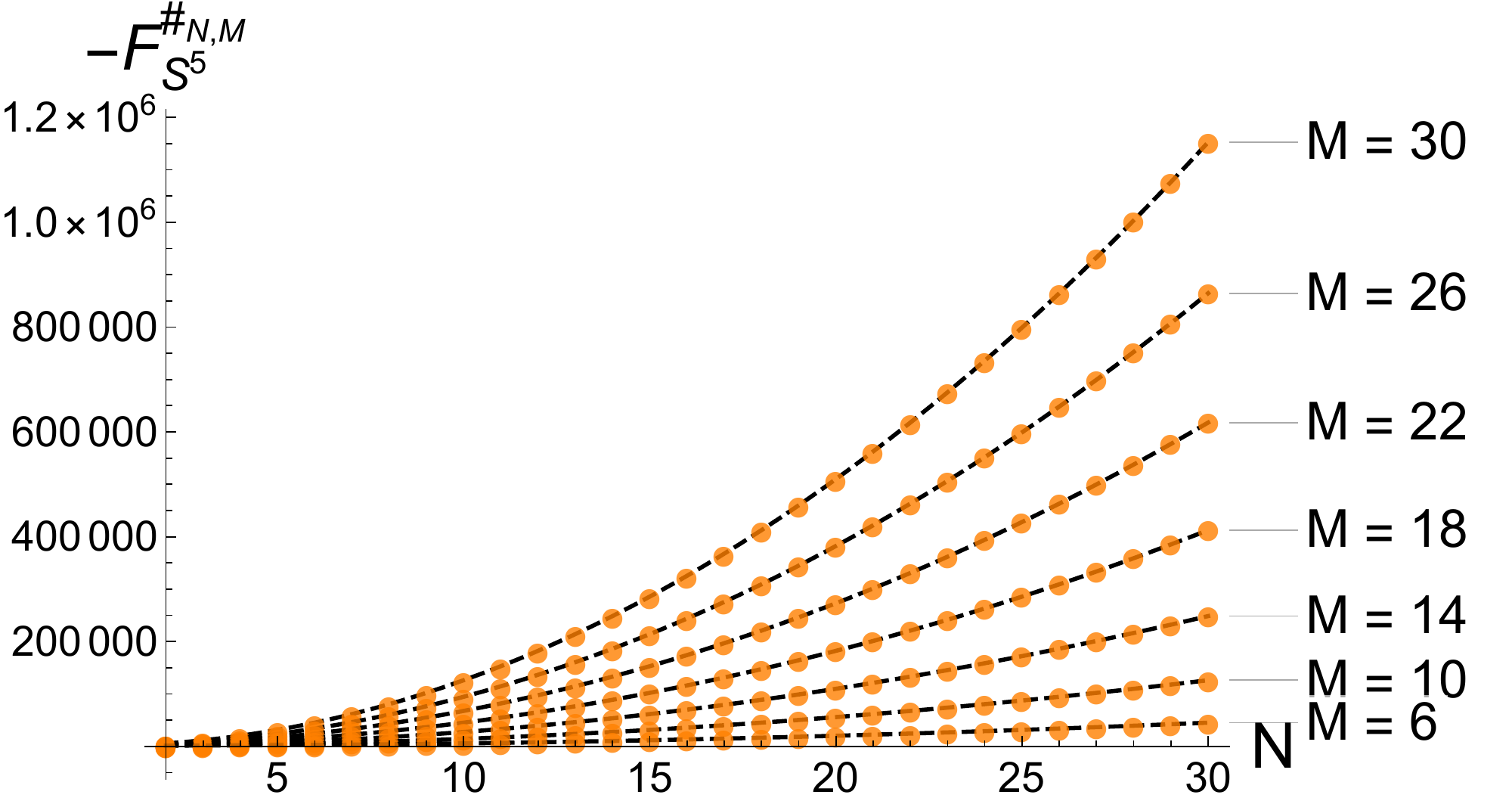}
 }\hskip -0.1in
 \subfloat{
  \includegraphics[width=0.49\linewidth]{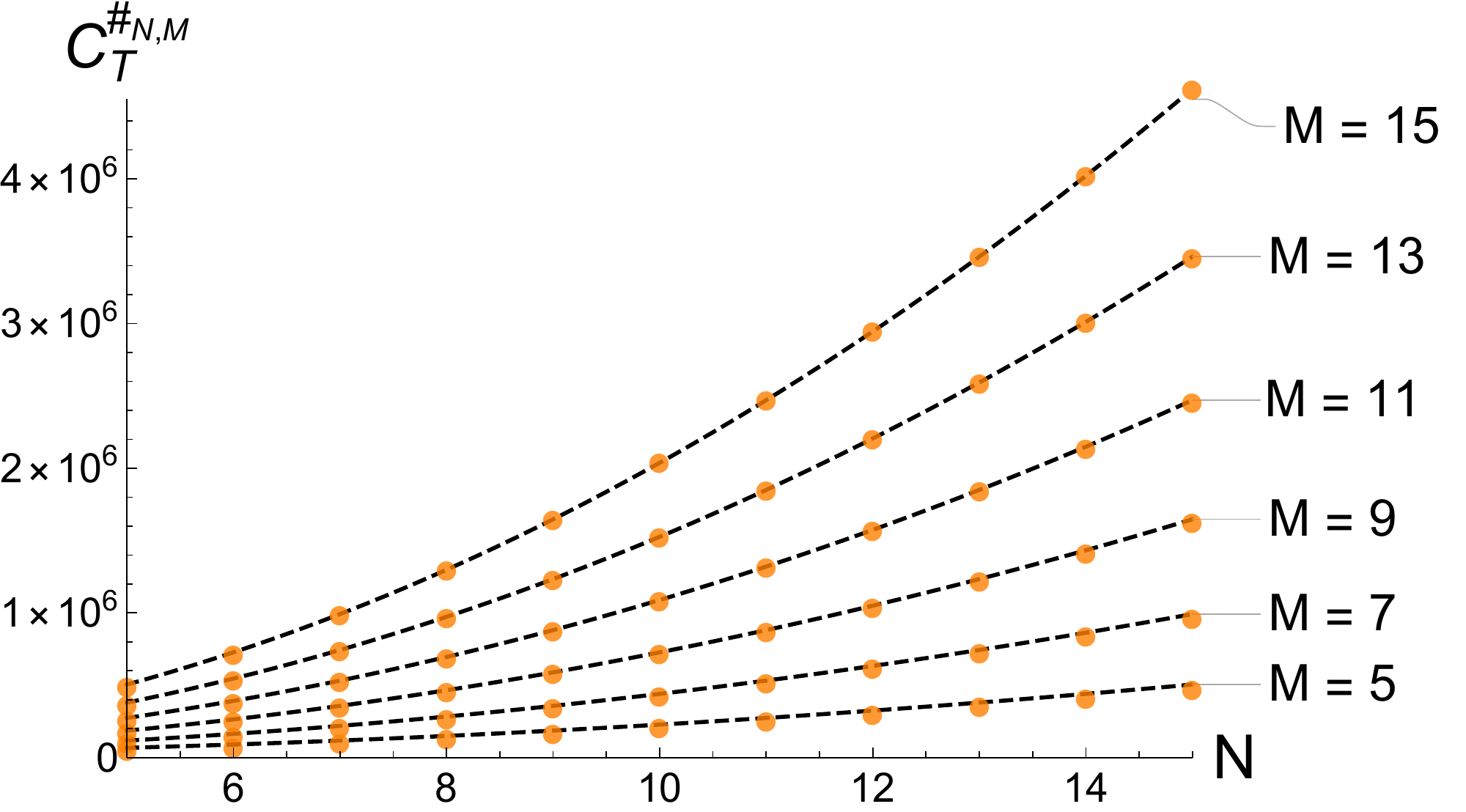}
 }
 \caption{The curves on the left hand side show the fit to the five-sphere partition functions $-F_{\rm S^{5}}^{\#_{N,M}}$ for the $\#_{N,M}$ theories,~(\ref{eq:hash-fit}) with~(\ref{eq:hash-fit-parms}), as functions of $N$ for a representative set of fixed $M$. The numerical results are shown as dots, and the fit functions as dashed curves. The agreement for the remaining values of $M$ is similarly precise. Analogous plots for $C^{\#_{N,M}}_T$ are shown on the right hand side.\label{fig:hash-fit}}
\end{figure}

These results indicate that, for large $N$ and $M$, the partition functions obtained from the quiver deformations indeed provide a good approximation to the superconformal field theory partition functions. For fixed $N$, the partition functions exhibit a clear quadratic scaling with $M$, and, likewise, for fixed $M$ they exhibit a clear quadratic scaling with $N$. The excellent agreement between the results obtained from the quiver~(\ref{eq:HMN-quiver}) and the S-dual quiver~(\ref{eq:HMN-dual-quiver}) for large $N$, $M$ motivates a fit to a quadratic polynomial in the $SL(2,\ZZ)$-invariant combination $(N M)$,
\begin{align}\label{eq:hash-fit}
 -F_{\rm fit}^{\#_{N,M}}& \ = \ \sum_{i=0}^2 a_i \left(N M\right)^i\,.
\end{align}
To quantitatively extract the large-$N,M$ behavior, we restrict the data to the subset~(\ref{eq:hash-restriction}),
and a least-squares fit results in
\begin{align}\label{eq:hash-fit-parms}
 a_2& \ = \ 1.43834\,, & a_1& \ = \ -15.7459\,, & a_0& \ = \ 1556.67\,.
\end{align}
The maximal relative error for this fit function is less than $1\%$, which occurs for $M$, $N$ at the lower end of~(\ref{eq:hash-restriction}), and the error decreases rapidly towards larger values of $N$, $M$. The quality of the fit is illustrated in Figure~\ref{fig:hash-fit}. We once again emphasize that the subleading terms may be subject to corrections, qualitatively and quantitatively, but that these corrections are irrelevant for the comparison to supergravity.

The discussion for the conformal central charge proceeds analogously. Explicit numerical results for $N\leq 15$ can be found in Appendix~\ref{App:Numerics}. There is again a clear quadratic scaling with $N$ for fixed $M$ and with $M$ for fixed $N$. 
Following the same logic as for the partition functions, we choose the fit function
\begin{align}
 C_{T,{\rm fit}}^{\#_{N,M}}& \ = \ \sum_{i=0}^2 b_i (N M)^i\,.
\end{align}
A least-squares fit, with $11\leq M, N\leq 15$, yields
\begin{align}
 b_2& \ = \ 93.4629\,, & b_1& \ = \ -597.753\,, & b_0& \ = \ 24288.3\,.
\end{align}
The maximal relative error is less than $1\%$, and occurs for maximal $|N-M|$. A graphical illustration of the fit quality can be found in Figure~\ref{fig:hash-fit}.

In summary, the behavior of the five-sphere partition function and conformal central charges in the limit of large $N$ for the $T_N$ theory, as extracted from supersymmetric localization, is given by
\begin{align}\label{eq:F-CT-TN-largeN}
 -F_{(N\gg 1)}^{T_N}& \ = \  0.411184\, N^4\,, & C_{T,(N\gg 1)}^{T_N}& \ = \  26.6647\,N^4\,,
\end{align}
while the behavior for large $N$, $M$ for the $\#_{N,M}$ theory is given by
\begin{align}\label{eq:F-CT-hash-largeN}
 -F_{(N,M\gg 1)}^{\#_{N,M}}& \ = \  1.43834\, N^2M^2\,, & C_{T,(N,M\gg 1)}^{\#_{N,M}}& \ = \  93.4629\, N^2M^2\,.
\end{align}

\section{Partition functions and central charges from supergravity}\label{sec:3}

In this section we review the supergravity solutions of~\cite{DHoker:2016ujz,DHoker:2017mds} and the results for the on-shell actions obtained in~\cite{Gutperle:2017tjo}. We then give explicit expressions for the five-sphere partition functions and conformal central charges for the $T_N$ and $\#_{N,M}$ theories obtained from holography.

\subsection{Solutions and on-shell action}

The supergravity solutions in~\cite{DHoker:2016ujz} have a geometry given by a warped product of AdS$_6\times {\rm S}^2$ over a Riemann surface $\Sigma$, and they involve non-trivial axion-dilaton and two-form fields. The entire solution is parametrized by two locally holomorphic functions $\cA_\pm$ on $\Sigma$. Physically regular solutions that are naturally associated to five-brane webs were constructed in~\cite{DHoker:2017mds}. For these solutions $\Sigma$ is taken as a disc or, equivalently, the upper half plane with complex coordinate $w$, on which the locally holomorphic functions take the form
\begin{align}\label{eq:cA}
 \cA_\pm & \ = \ \cA_\pm^0+\sum_{\ell=1}^L Z_\pm^\ell \ln(w-p_\ell)\,, 
 &
 \sum_{\ell=1}^L Z_\pm^\ell  & \ = \ 0\,,
 &
 Z_\pm^\ell & \ = \ -\bar Z_\mp^\ell\,,
\end{align}
with complex constants $Z_\pm^\ell$ that are discussed in detail in \cite{DHoker:2017mds}. The supergravity fields are conveniently expressed in terms of the composite quantities
\begin{align}\label{eq:composites}
 \kappa ^2 & \ = \  - |\partial_w \cA_+|^2 + |\partial_w \cA_-|^2\,, & 
 \partial_w \cB & \ = \  \cA_+ \partial_w \cA_- - \cA_- \partial_w \cA_+\,, \no\\
 \cG &  \ = \   |\cA_+|^2 - |\cA_-|^2 + \cB + \bar \cB\,,&
 R+\frac{1}{R} & \ = \   2+ { 6 \, \kappa^2 \, \cG \over |\partial_w\cG|^2}\,.
\end{align}
The metric and complex two-form field $C_{(2)}$ are parametrized as
\begin{align}
 \diff s^2 &  \ = \    f_6^2 \, \diff s^2 _{{\rm AdS}_6} + f_2 ^2 \, \diff s^2 _{{\rm S}^2} +  4 \rho^2 \left| \diff w \right|^2\,,
 &
 C_{(2)}& \ = \ \cC \vol_{\rm S^2}\,,
\end{align}
where the functions on $\Sigma$ appearing in the metric are given by
\begin{align}
f_6^2 & \ =  \ \sqrt{6\cG} \left ( \frac{1+R}{1-R} \right ) ^{\tfrac{1}{2}},
&
f_2^2 & \ = \  \frac{1}{9}\sqrt{6\cG} \left ( \frac{1-R}{1+R} \right )^{\tfrac{3}{2}},
&
\rho^2 \ & = \ {\kappa^2 \over \sqrt{6\cG} } \left (\frac{1+R}{1-R} \right )^{\tfrac{1}{2}},
\end{align}
and the function $\cC$ parametrizing the complex two-form field is given by
\begin{align}\label{eqn:flux}
 \cC  \ = \  \frac{4 \ii }{9}\left (  
\frac{\partial_{\bar w} \bar \cA_- \, \partial_w \cG}{\kappa ^2} 
- 2 R \, \frac{  \partial_w \cG \, \partial_{\bar w} \bar \cA_- +  \partial_{\bar w}  \cG \, \partial_w \cA_+}{(R+1)^2 \, \kappa^2 }  
 - \bar  \cA_- - 2 \cA_+ \right )\,.
\end{align}
The axion-dilaton scalar $B=(1+\ii \tau)/(1-\ii\tau)$ is given by 
\begin{align}\label{eq:B}
B & \ = \ \frac{\partial_w \cA_+ \,  \partial_{\bar w} \cG - R \, \partial_{\bar w} \bar \cA_-   \partial_w \cG}{
R \, \partial_{\bar w}  \bar \cA_+ \partial_w \cG - \partial_w \cA_- \partial_{\bar w}  \cG}\,.
\end{align}
The differentials $\partial_w\cA_\pm$ resulting from~(\ref{eq:cA}) have $L\geq 3$ poles on the real line with residues given by $Z_\pm^\ell$. At these poles, the external $(p,q)$ five-branes of the associated brane web emerge, with their $(p,q)$-charges determined by the residues. For a given choice of residues, subject to the constraints in~(\ref{eq:cA}), the remaining parameters are determined by regularity conditions, which are discussed in detail in~\cite{DHoker:2017mds}.

Since the five-form field vanishes in this class of solutions, the Type IIB supergravity on-shell action can be evaluated straightforwardly. For Euclidean AdS$_6$ in global coordinates, such that the dual superconformal field theory is defined on $\rm S^5$, the general result is \cite{Gutperle:2017tjo}
\ie\label{eq:IIBaction}
 S_{\rm IIB}& \ = \ \frac{1}{8\pi G_N}\Vol_{{\rm AdS}_6,{\rm ren}}\Vol_{{\rm S}^2}\,I_0\,,
\fe
where $\Vol_{{\rm AdS}_6,{\rm ren}}$ is the renormalized unit-radius AdS$_6$ volume,
\begin{align}
 \Vol_{{\rm AdS}_6,{\rm ren}}& \ = \ -\frac{8}{15}\Vol_{\rm S^5}\,,
 &
 \Vol_{{\rm S}^5}& \ = \ \pi^3\,,
 &
 \Vol_{{\rm S}^2}& \ = \ 4\pi\,.
\end{align}
For a general solution with three poles (equation~(52) of~\cite{Gutperle:2017tjo})
\ie\label{I0-3pole}
I_0& \ = \ -80\pi \zeta(3) \left(Z_+^1 Z_-^2 - Z_+^2 Z_-^1\right)^2\,, 
\fe
while for a solution with four poles and pairwise opposite-equal residues (equation~(55) of~\cite{Gutperle:2017tjo})
\ie\label{I0-4pole}
 I_0& \ = \ -280\pi\zeta(3)\left(Z_+^1 Z_-^2 - Z_+^2 Z_-^1\right)^2\,.
\fe

\subsection{\texorpdfstring{$T_N$ and $\#_{N,M}$}{T-N and \#-N-M} partition functions}

We now discuss the supergravity solutions that are naturally associated with the brane webs of Figure~\ref{fig:webs} in the conformal limit, and obtain the holographic results for the five-sphere partition functions.
The precise relation of the residues $Z_\pm^\ell$ to the $(p,q)$ five-brane charges is obtained from the charge quantization conditions. From Section~3.9 of~\cite{DHoker:2017mds}, the complex three-form field strength near the pole $p_m$ is given by
\ie
\diff C_{(2)}& \ \approx \ \frac{8}{3}Z_+^m \vol_{\rm S^3}\,,
\fe
where ${\rm S}^3$ denotes the three-sphere formed in the geometry around the pole. 
With the NSNS two-form $B_2$ and the RR two-form $C_{(2)}^{\rm RR}$, $C_{(2)}=B_2+\ii C_{(2)}^{\rm RR}$.
We follow the conventions of~\cite{deAlwis:1997gq} for the normalization of the effective actions.
The Dirac quantization conditions yield
\begin{align}
 T\int_{{\rm S}^3} \diff (B_2+\ii C_{(2)}^{\rm RR}) & \ = \ 2\pi (q+\ii p)\,,&
 T & \ = \ \frac{1}{2\pi\alpha^\prime}\,,
\end{align}
and the identification of the five-brane charges with the residues as
\ie\label{eq:residue-N}
 Z_+^m& \ = \ \frac{3}{4}\alpha^\prime \left(p+\ii q\right)\,.
\fe
To explicitly evaluate the Type IIB supergravity on-shell action, we also need the ten-dimensional Newton constant, given by
\ie\label{eq:Newton10d}
 2\kappa^2_{10}& \ = \ 16\pi G_N \ = \ (2\pi)^7{\alpha^\prime}^4\,.
\fe

The solution for the $T_N$ theory, which corresponds to a junction of $N$ D5-branes, $N$ NS5-branes and $N$ (1,1) five-branes, is consequently realized by a three-pole solution, $L=3$, with residues
\begin{align}\label{eq:N-junction-residues}
 Z_+^1& \ = \ \frac{3}{4}\alpha^\prime N\,,\quad
 &Z_+^2& \ = \ \frac{3}{4}\ii\alpha^\prime N\,,\quad
 &Z_+^3& \ = \ -\frac{3}{4}\alpha^\prime(1+\ii)N\,.
\end{align}
A regular solution is obtained by placing the poles at $p_1=1$, $p_2=0$, $p_3=-1$ and fixing 
\ie
\cA_+^0 \ = \ -\bar\cA_-^0 \ = \ \frac{3\ii}{4}\alpha^\prime  N\ln 2~.
\fe

The solution for the $\#_{N,M}$ theory, corresponding to an intersection of $N$ D5-branes and $M$ NS5-branes, is realized by a four-pole solution with residues
\begin{align}\label{eq:D5-NS5-residues}
 -Z_+^1\ = \ Z_+^3& \ = \ \frac{3}{4} \ii \alpha^\prime N\,, \quad &
 Z_+^2 \ = \ -Z_+^4& \ = \ \frac{3}{4}\alpha^\prime M\,.
\end{align}
The regularity conditions are solved by fixing $p_1=1$, $p_2=2/3$, $p_3=1/2$, $p_4=0$ and $\cA_+^0=-\bar\cA_-^0=Z_+^2\ln 3-Z_+^1\ln 2$.

Using the on-shell action~(\ref{eq:IIBaction}) with~(\ref{I0-3pole}) and the residues in~(\ref{eq:N-junction-residues}) yields the on-shell action for the $T_N$ solution as
\ie\label{eqn:STNsugra}
 S_{\rm IIB}(T_N)& \ = \ -\frac{27}{8\pi^2}\,\zeta(3)N^4\,.
\fe
Similarly, using~(\ref{I0-4pole}) and the residues in~(\ref{eq:D5-NS5-residues}) yields the on-shell action for the $\#_{N,M}$ solution as
\ie\label{eqn:SHTsugra}
 S_{\rm IIB}(\#_{N,M})& \ = \ -\frac{189}{16\pi^2}\,\zeta(3)N^2M^2\,.
\fe
The five-sphere partition functions of the dual superconformal field theories, with the sign conventions for the Euclidean action as in~\cite{Gutperle:2017tjo}, are given by $F=S_{\rm IIB}$.

\subsection{Conformal central charges}\label{sec:CT-IIB}

We now discuss the conformal central charge $C_T$, defined in Appendix \ref{App:CT}. As shown in~\cite{Gutperle:2018wuk}, $C_T$ is related to the five-sphere partition function by a simple rescaling for the superconformal field theories described by the warped ${\rm AdS}_6$ solutions in Type IIB supergravity. The ratio can be obtained explicitly as follows. The effective six-dimensional gravitational coupling $\kappa_6$ is given by \cite{Gutperle:2018wuk}
\begin{align}
 \frac{1}{\kappa_6^2} & \ = \ \frac{8}{3\kappa_{10}^2}\Vol_{{\rm S}^2}\int \diff^2w\,\kappa^2\cG\,.
\end{align}
With $C_T$ as given \cite{Freedman:1998tz,Penedones:2016voo} (see also Section~4.1 of~\cite{Chang:2017mxc} with $\ell=1$ and $d=5$), this yields
\ie
 C_T \ = \ \frac{5\cdot 2^{13}\pi}{9\kappa_{10}^2}\Vol_{\rm S^2}\int \diff^2w\, \kappa^2\cG\,.
\fe
As shown in~\cite{Gutperle:2017tjo}, the five-sphere partition function is equal to the finite part of the entanglement entropy for a spherical region for the warped AdS$_6$ Type IIB solutions. This entanglement entropy in turn is given by (section~IV of~\cite{Gutperle:2017tjo})
\ie
 S_{\rm EE,finite} & \ = \ \frac{64}{9\kappa_{10}^2}\pi^3\Vol_{{\rm S}^2}\int \diff^2 w\,\kappa^2\cG\,.
\fe
We therefore find 
\ie
 C_T& \ = \ - \frac{640}{\pi^2}F_{{\rm S}^5}\,.
\fe
With the result in~\eqref{eqn:STNsugra}, the supergravity prediction for the conformal central charge for $T_N$ theories is then
\ie\label{eq:CT-TN-sugra}
C_{T}^{T_N} & \ = \ \frac{2160}{\pi^4}\,\zeta(3)N^4\,,
\fe
and similarly equation~\eqref{eqn:SHTsugra} gives the following central charge for the $\#_{N,M}$ theories from supergravity 
\ie\label{eq:CT-hash-sugra}
C_{T}^{\#_{N,M}} & \ = \ \frac{7560}{\pi^4}\,\zeta(3)N^2M^2\,.
\fe

This provides another non-trivial prediction to be matched by the field theory computations at large $N$. The relation is the same as found for the Type IIA supergravity solutions in~\cite{Chang:2017mxc}, suggesting that it holds more generally for holographic conformal field theories. In Appendix~\ref{sec:app-CT} we show that it can indeed be derived with rather generic assumptions on the Kaluza-Klein reduced effective action.

\section{Comparison}
\label{sec:comp}

In this section we compare the field theory results for $F_{\rm S^{5}}$ and $C_T$ for the $T_N$ and $\#_{N,M}$ five-dimensional superconformal field theories obtained in Section~\ref{sec:largeN} to the corresponding supergravity results of Section~\ref{sec:3}. 

\begin{figure}
\centering
\subfloat{
\includegraphics[width=.48\textwidth]{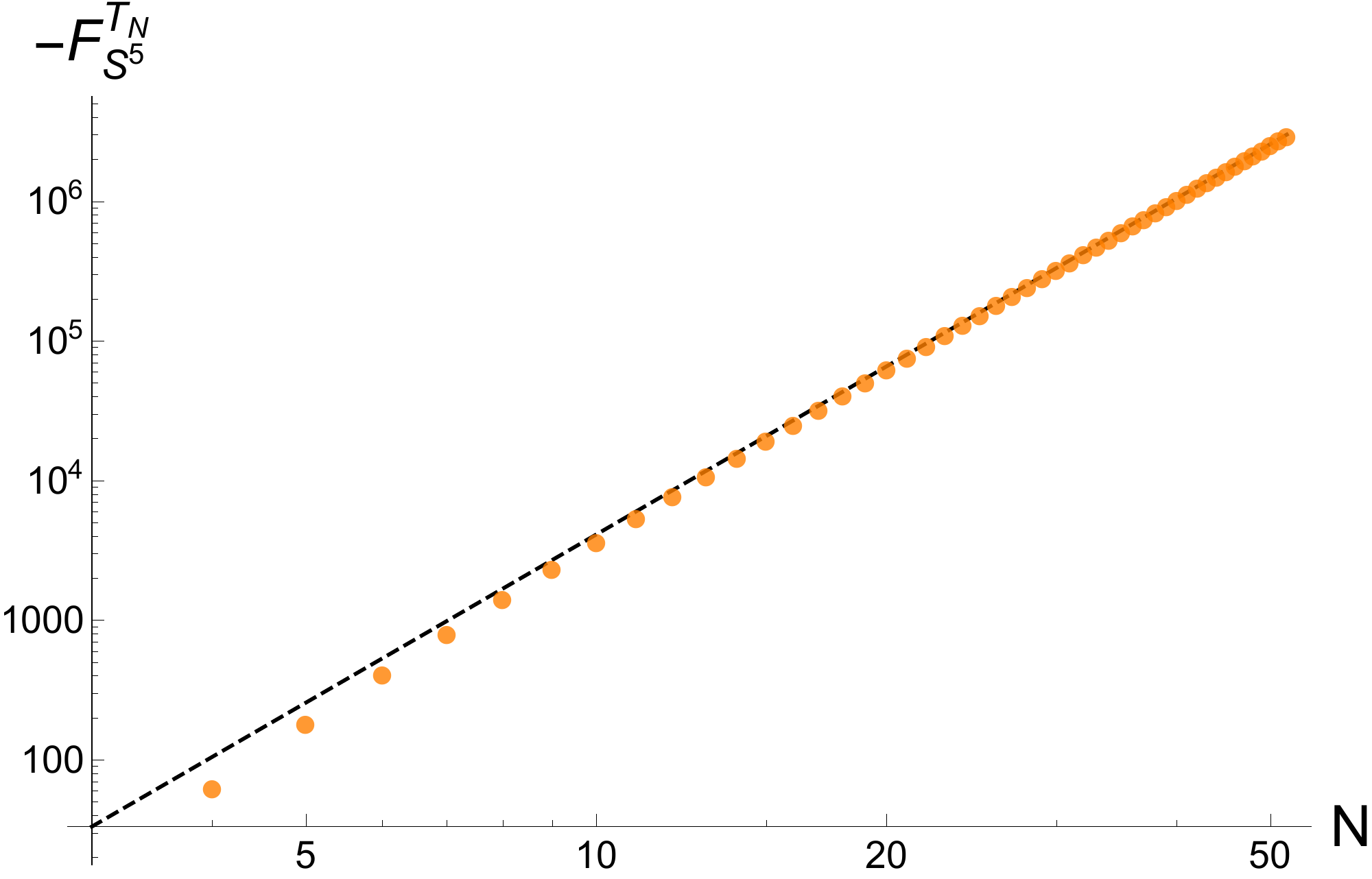}
}
\hskip 0.0in
\subfloat{
\includegraphics[width=.48\textwidth]{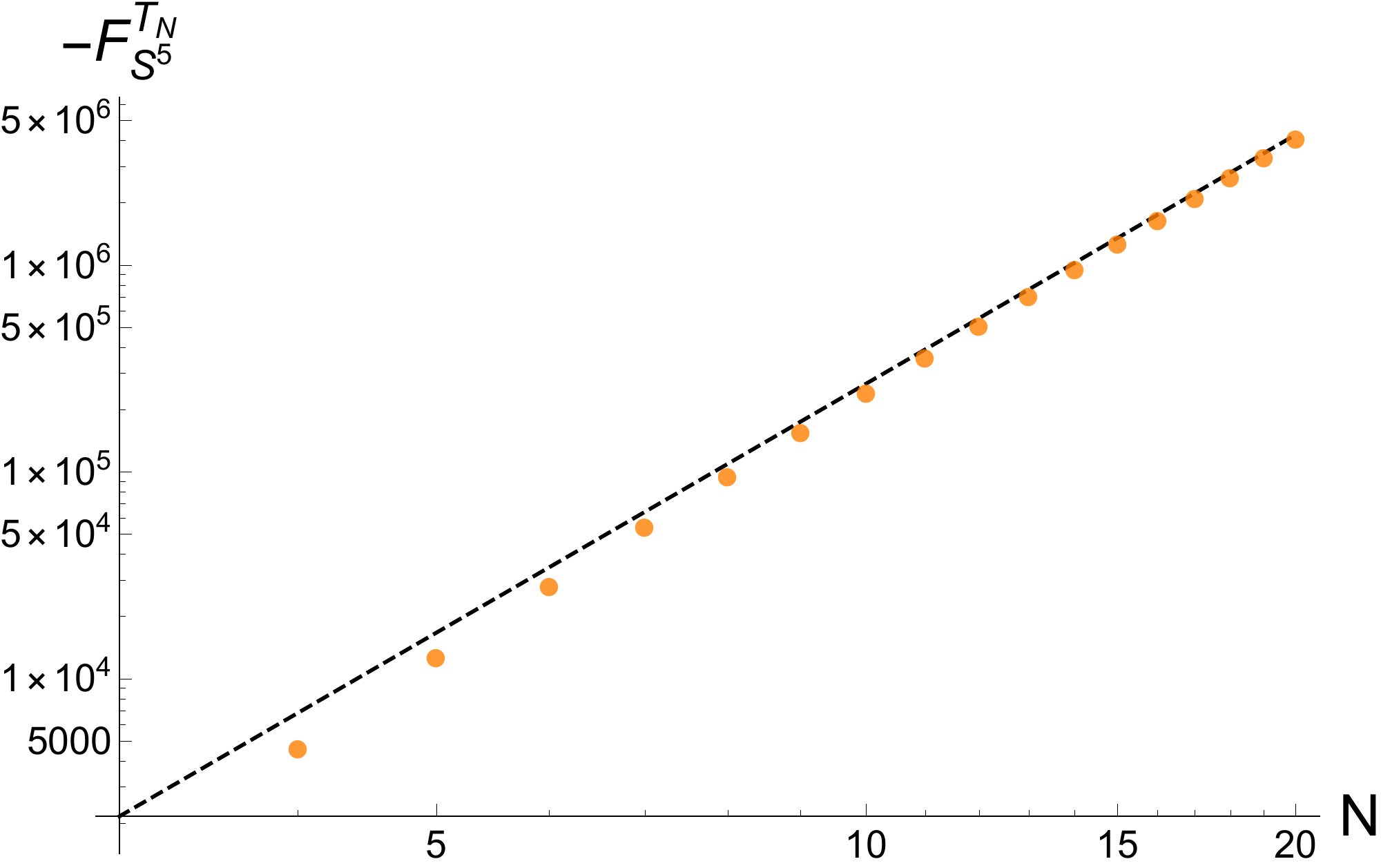}
}
\caption{On the left hand side the numerically obtained finite-$N$ results for the five-sphere partition function of the $T_N$ theories as orange points, and the supergravity result as dashed line. On the right hand side the analogous plot for $C_T$.}
\label{Fig:LogLogFvsSugraTN}
\end{figure}

The large $N$ behavior of the five-sphere partition function and the conformal central charge for the $T_N$ theory, as obtained from supersymmetric localization, were given in~(\ref{eq:F-CT-TN-largeN}), while the corresponding supergravity results are in~(\ref{eqn:STNsugra}) and~(\ref{eq:CT-TN-sugra}), respectively. The first crucial check is for the scaling at large $N$. Both computations agree on the quartic scaling with $N$, for both quantities. For the overall numerical coefficient, we find
\begin{align}
  F_{(N\gg 1)}^{T_N}& \ = \  1.00032\, F_{{\rm sugra}}^{T_N}\,, &
 C_{T,(N\gg 1)}^{T_N}& \ = \  1.00036\,C_{T,{\rm sugra}}^{T_N}\,.
\end{align}
This shows that the large $N$ behavior of the field theory partition function agrees to remarkable accuracy with the supergravity result. We recall that the fit function from which the large $N$ behavior of the field theory results was extracted in Section~\ref{sec:num-results} reproduces the actual numerical data with relative errors of at most $\mathcal O(10^{-3})$ for all $N>10$. This is roughly the order to which we find agreement of the large $N$ behavior. Plots comparing the field theory results for finite $N$ to the supergravity predictions are shown in Figure~\ref{Fig:LogLogFvsSugraTN}. They show that the scaling as well as the coefficient of the leading term agree to high accuracy between the supergravity and field theory computations. These results certainly support the identification of the three-pole supergravity solution as the holographic dual for the $T_N$ theory.

\begin{figure}
 \centering
 \subfloat{
  \includegraphics[width=0.49\linewidth]{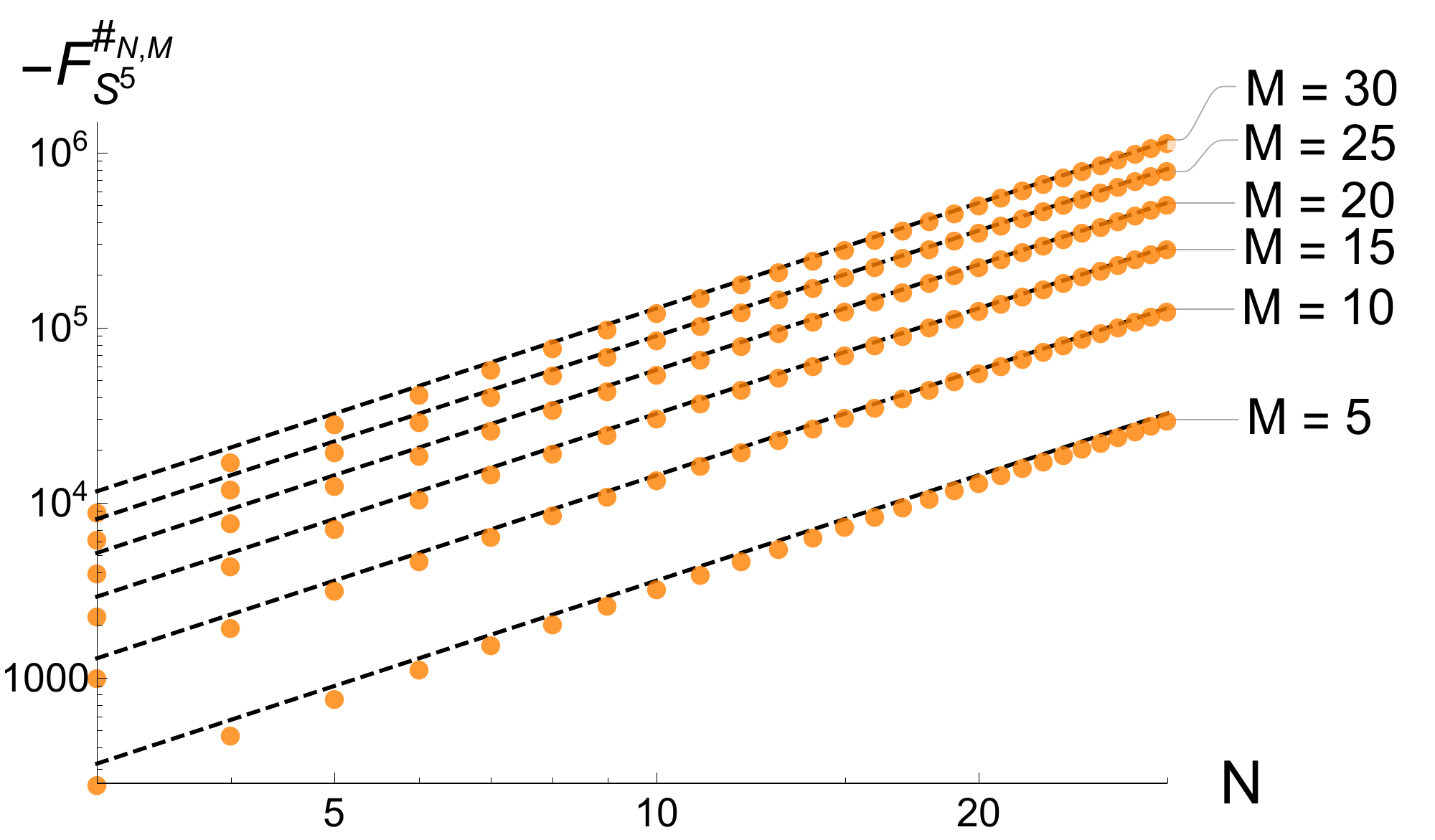}
 }\hskip - 0.1in
 \subfloat{
  \includegraphics[width=0.49\linewidth]{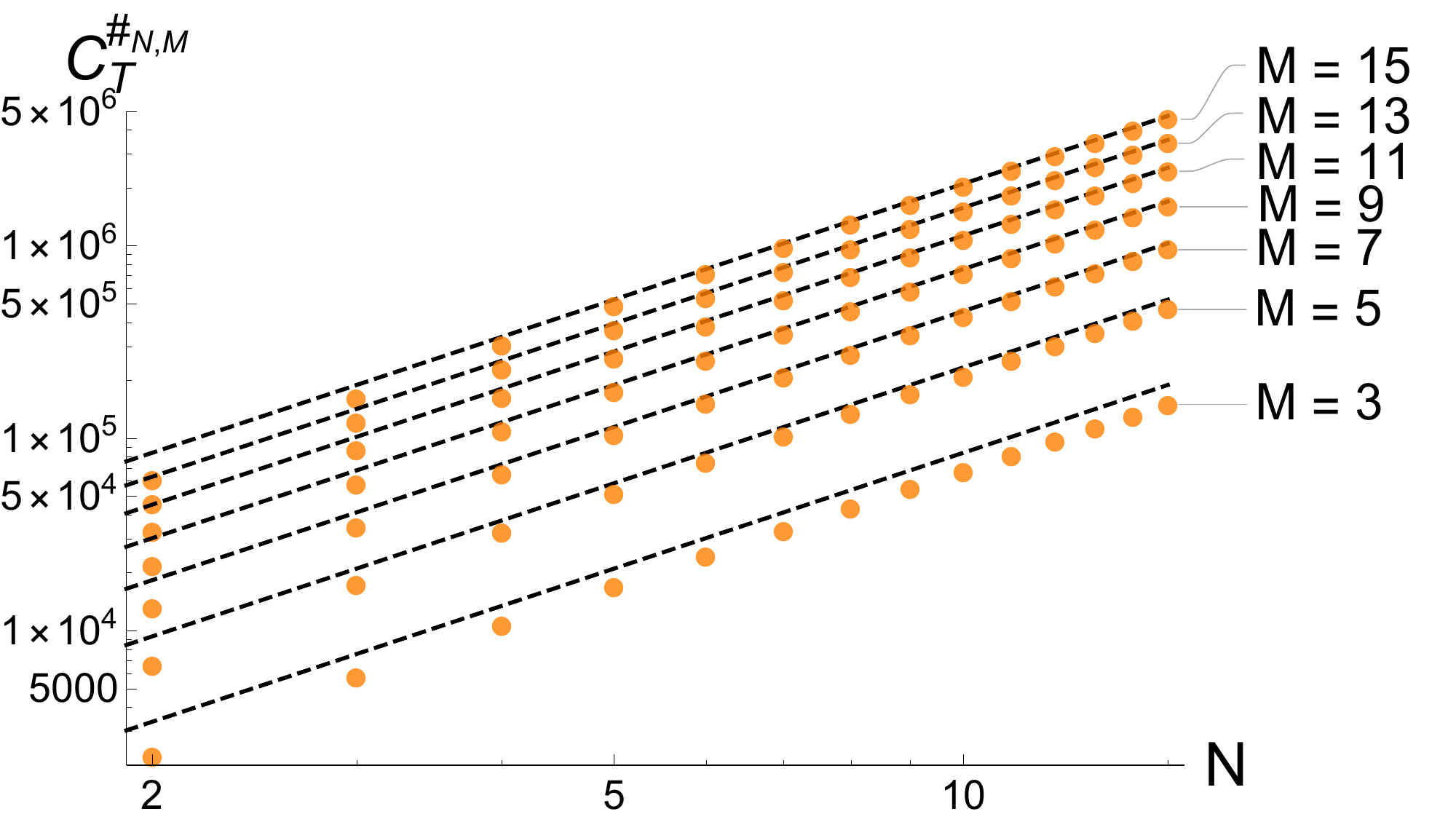}
 }
\caption{On the left hand side the results for $-F_{\rm S^{5}}^{\#_{N,M}}$ discussed in Section~\ref{sec:num-results} for a selection of fixed $M$. The dashed lines show the supergravity results for the corresponding fixed $M$. The plots for the remaining values of $M$ similarly show precise agreement. On the right hand side we show the same plot for the conformal central charge.}
\label{Fig:LogLogFvsSugraM}
\end{figure}

For the $\#_{N,M}$ theory the behavior of the sphere partition function and conformal central charge at large $N$ and $M$, as obtained from localization, were given in~(\ref{eq:F-CT-hash-largeN}), with the corresponding supergravity results in \eqref{eqn:SHTsugra} and~(\ref{eq:CT-hash-sugra}), respectively. Both computations agree on the scaling $N^2M^2$ for both quantities. The scaling with $N^2$ coincides with the scaling of the partition function in the quiver deformation~(\ref{eq:HMN-quiver}) in the limit where the theory is free. Such a weak coupling analysis, however, would yield a linear scaling with $M$, in contrast to the $M^2$ scaling exhibited by the actual results. The $M^2$ scaling thus is a genuine strong coupling effect in the quiver~(\ref{eq:HMN-quiver}). In the dual quiver,~(\ref{eq:HMN-dual-quiver}), the roles of $N$ and $M$ are, naturally, reversed. For the overall numerical coefficients we obtain the following relation of localization and supergravity results,
\begin{align}
 F_{{\rm sugra}}^{\#_{N,M}}& \ = \  0.999758\, F_{(N\gg 1)}^{\#_{N,M}}\,, &
 C_{T,(N\gg 1)}^{\#_{N,M}}& \ = \  1.00183\,C_{T,{\rm sugra}}^{\#_{N,M}}\,.
\end{align}
This once again shows remarkable agreement between the field theory and supergravity computations. Plots comparing the field theory results for finite $N$ to the supergravity prediction can be found in Figure~\ref{Fig:LogLogFvsSugraM}. The accurate agreement of supergravity and field theory computations supports the identification of the proposed supergravity dual with the superconformal field theory also in this case.

\section{Discussion}\label{sec:4}

In this paper we have initiated a detailed comparison of the Type IIB supergravity solutions of \cite{DHoker:2016ysh,DHoker:2017mds} to the putative dual superconformal field theories at the level of the partition functions. Gauge theory deformations of the dual superconformal field theories are typically given by quiver gauge theories with two large parameters: the maximal rank of the involved gauge groups and the number of nodes. This is in particular the case for the two theories we have studied here, namely the $T_N$ theories and the $\#_{N,M}$ theories, realized on an intersection of D5- and NS5-branes. The large Coulomb branches complicate analytic computations of the partition functions using supersymmetric localization. We have therefore employed numerical methods, and considered the field theories at finite $N$, up to $N=52$, for the $T_N$ theories and finite $N$, $M$, up to $M=N=30$, for the $\#_{N,M}$ theories. From this large sample of explicit data we then extracted the scaling of the five-sphere partition functions and conformal central charges $C_T$ at large $N$ and large $N,M$, respectively. The results accurately match the corresponding quantities obtained from the proposed supergravity duals for both theories, in scaling and in the leading coefficient. This lends strong support to the identification of the supergravity solutions with the proposed field theories.

For the finite $N$ and $(N,M)$ computations we have employed the approximations that are typically used in large $N$ limits. Namely, we dropped instanton contributions to the partition functions and used saddle point techniques. Both of these approximations are expected to become exact in the appropriate large $N$ and large $(N,M)$ limits, which is the part of interest for the comparison to supergravity. For finite $N$ and $(N,M)$ we have estimated the impact of these approximations by comparing the $\#_{N,M}$ quiver gauge theory to the dual quiver with $N$ and $M$ exchanged. This comparison showed remarkable agreement already for moderate values of $N$ and $M$, with the discrepancy vanishing with increasing $N$, $M$. This is in line with general expectations and supports the use of these approximations.

The results presented here are compatible with a potential $F$-theorem, and perhaps even a $C_T$-theorem, in five dimensions: To the extent that the approximations discussed above do not qualitatively distort the finite $N$ picture, the free energies and conformal central charges obtained here are consistent with possible flows between the theories and monotonicity of $F$ and $C_T$. This is in line with previous observations to the same effect in~\cite{Klebanov:2011gs,Jafferis:2012iv,Chang:2017cdx,Chang:2017mxc,Gutperle:2017tjo}. In the appropriate large $N$ limits, the field theory results in particular exhibit the universal relation between $F_{\rm S^5}$ and $C_T$ realized in holographic theories (discussed in Appendix~\ref{sec:app-CT}). Due to this relation, an $F$-theorem is equivalent to a $C_T$-theorem for theories which are holographic in the sense that they satisfy this relation. At finite $N$, however, the two quantities are independent, and may or may not satisfy monotonicity theorems individually.

An obvious direction for future research is the analytic evaluation of the five-sphere partition functions and central charges. The explicit numerical results, including the eigenvalue distributions, provide useful guidance for analytic computations, which we are currently investigating. Aside from analytic results, there are several novel ways in which the AdS$_{6}$/CFT$_{5}$ correspondences in Type IIB string theory can be tested using methods similar to the ones employed here. 
First of all, there are additional classes of five-dimensional superconformal field theories that are conjectured to be captured by the general Type IIB solutions, e.g.\ theories realized by five-brane webs with additional seven-branes \cite{DHoker:2017zwj}. It would be interesting to provide explicit checks also of these proposed dualities.
Furthermore, we can extend the checks of $T_N$ and $\#_{N,M}$ to include flavor central charges (see~\cite{Chang:2017mxc} for such checks in massive IIA), which provide detailed probes not only of the supergravity solutions but also of flavor symmetry enhancement of the five-dimensional superconformal field theories. Other potential checks include squashed five-sphere or squashed Sasaki-Einstein partition functions, as well as Wilson loop expectation values (see~\cite{Assel:2012nf,Alday:2015jsa,Alday:2014bta,Alday:2014rxa} for such checks in massive IIA AdS$_{6}$ solutions).

\section*{Acknowledgements}

We thank Diego Rodr\'iguez-G\'omez for helpful discussions and comments. MF is supported by the David and Ellen Lee Postdoctoral Scholarship and the U.S. Department of Energy, Office of Science, Office of High Energy Physics, under Award Number DE-SC0011632. CFU is supported in part by the National Science Foundation under grant PHY- 16-19926.

\clearpage
\appendix

\section{Squashed five-sphere partition functions and conformal central charges}
\label{App:PartFunc}

In this section we first briefly review the general form of the perturbative part of the (squashed) five-sphere partition function as derived in the references~\cite{Kallen:2012va,Kim:2012ava,Imamura:2012bm,Lockhart:2012vp,Imamura:2012xg}. We shall then provide some generalities about the triple-sine functions appearing in the partition function in~Appendix~\ref{app:tripsine}, and outline the relation of the conformal central charge $C_T$ to squashing deformations of the five-sphere partition function as derived in~\cite{Chang:2017cdx} in~Appendix~\ref{App:CT}. The relevant formulas are used in the main text to compute the round sphere partition function as well as the conformal central charges of the ultraviolet superconformal field theories.

The metric on a squashed (unit) five-sphere of $\rm U(1)\times U(1)\times U(1)$ isometry can be explicitly written as
\ie\label{eqn:sqmet}
\diff s^{2} \ = \ \sum_{i=1}^{3}\left( \diff y_i^{2} + y_{i}^{2} \diff \phi_{i}^{2} \right) + \tilde\kappa^{2} \left( \sum_{j=1}^{3}a_j y_j^{2}\diff \phi_{j} \right) ^{2} \,,\qquad \tilde \kappa^{2} \ = \ \frac{1}{1-\sum_{j=1}^{3}y_{j}^{2}a_{j}^{2}} \,,
\fe
with periodic coordinates $\phi_j \sim \phi_j + 2\pi$. The first summand is the round (unit) five-sphere, and thus the squashed metric can be viewed as a deformation of the round sphere metric by turning on real non-vanishing squashing parameters $a_i \neq 0$. 

The perturbative part of the squashed five-sphere of $\rm U(1) \times SU(3) $ isometry has been derived from a reduction of the six-dimensional superconformal index, and conjectured for general squashing in~\cite{Imamura:2012xg}. For a five-dimensional gauge theory of gauge group $G$, with $N_f$ hypermultiplets in a (real) representation $R_f \otimes \bar R_{f}$ of $G$, the perturbative part of the squashed five-sphere partition function is explicitly given by
\ie\label{eqn:generalpartfunc}
\cZ_{\rm pert} [\omega_1, \omega_2, \omega_3]  \ = \ & 
\frac{S_{3}^{\prime} \left( 0\mid \omega_1,\omega_2,\omega_3 \right)^{\mathrm{rk} \, G}}{\left| \cW \right| (2\pi)^{\mathrm{rk} \, G}} 
\left[ \prod_{i=1}^{\mathrm{rk} \, G} \int_{-\infty}^{\infty} \diff \lambda_i \right]e^{-\frac{(2\pi)^{3}}{\omega_1\omega_2\omega_3}\mathfrak{F}(\lambda)}\\
& 
\hspace{2.3 cm} \times 
\frac{\prod_{\alpha} S_3 \left( -\ii \alpha(\lambda) \mid \omega_1,\omega_2,\omega_3 \right)}{\prod_{f=1}^{N_f} \prod_{\rho_{f}} S_{3} \left( \ii \rho_{f}(\lambda) + \tfrac{\omega_{\rm tot}}{2} \mid \omega_1,\omega_2,\omega_3 \right)} \,,
\fe
where the products are taken over all roots $\alpha$ of $G$, the flavor hypermultiplets $f = 1, \ldots N_f$, as well as the weights $\rho_f$ of the corresponding representation $R_{f}\otimes \bar R_{f}$. Furthermore, we denote by $S_3 (z \mid \omega_1, \omega_2, \omega_3)$ the triple sine function, which we define in Section~\ref{app:tripsine}, with $\omega_i$ the squashing parameter related to $a_i$ in the metric~\eqref{eqn:sqmet} by
\ie
\omega_i \ = \ 1+ a_i \,,
\fe
and we additionally defined
\ie
\omega_{\rm tot} \ = \ \omega_1 + \omega_2 + \omega_3 \,.
\fe
Furthermore, we denote by $\left| \cW \right|$ the cardinality of the Weyl group $\cW$ of the gauge group $G$, and by $\mathfrak{F}(\lambda)$ the classical flat space prepotential, explicitly given by
\ie
\mathfrak{F} (\lambda ) \ = \ \frac{1}{2 g_{\rm YM}^{2}} \mathrm{Tr} \, \lambda^{2} + \frac{k}{6} \mathrm{Tr} \, \lambda^{3} \,,
\fe
with $g_{\rm YM}$ the classical five-dimensional gauge coupling, $k$ the Chern-Simons coupling and $\mathrm{Tr} (\lambda)$ the Killing form of $G$.

\subsection{Triple sine function}
\label{app:tripsine}

We now briefly review the definition of the triple sine function, $S_3 \left( z \mid \omega_1, \omega_2, \omega_3 \right)$, as it appears in the squashed five-sphere partition function of five-dimensional gauge theories. It can be defined as~\cite{Faddeev:1995nb,Ruijsenaars:2000,2003math......6164N,Faddeev:2012zu}
\bea
S_3 \left( z \mid \omega_1, \omega_2, \omega_3 \right) \ = \ \exp \left( 
- \frac{\pi \ii}{6} B_{3,3} \left( z \mid \omega_1, \omega_2, \omega_3 \right) 
- \cI_{3} \left( z \mid \omega_1, \omega_2, \omega_3 \right)\right) \,,
\eea
where one can show that $\cI_{3} \left( z \mid \omega_1, \omega_2, \omega_3 \right)$ has the following integral representation
\bea
\cI_{3} \left( z \mid \omega_1, \omega_2, \omega_3 \right) \ = \ \int_{\mathbb{R}+ \ii 0^{+}} \frac{\diff x}{x} \frac{e^{zx}}{ \left( e^{\omega_1 x} -1 \right)\left( e^{\omega_2 x} -1 \right)\left( e^{\omega_3 x} -1 \right)} \,,
\eea
with the contour of integration given by the real axis minus the origin at $x=0$, which is encircled by a semi-circle into the positive half-plane. Additionally, $B_{3,3} (z \mid \omega_1, \omega_2, \omega_3)$ is a generalized Bernoulli polynomial, explicitly given by
\bea
B_{3,3} (z \mid \omega_1, \omega_2, \omega_3) &\ = \ & 
\frac{1}{\omega_1\omega_2\omega_3} \bigg[ 
z^{3} 
-\frac{3}{2} \omega_{\rm tot} z^{2} 
+ \frac{1}{2} \left( \omega_{\rm tot}^{2}
+ \omega_1\omega_2 + \omega_2 \omega_3 + \omega_1 \omega_3 \right) \, z \nn\\
&& \qquad \qquad  - \frac{1}{4}\omega_{\rm tot} (\omega_1 \omega_2 + \omega_2 \omega_3 + \omega_1 \omega_3) \bigg] \,.
\eea
Using these formulae one can explicitly compute
\bea
S_3^{\prime} \left( 0 \mid 1, 1, 1 \right) \ = \  2 \pi  \exp \left( {\frac{\zeta (3)}{4 \pi ^2}} \right) \,.
\eea

\subsection{Conformal central charge \texorpdfstring{$C_T$}{C-T} from partition function}
\label{App:CT}
In references~\cite{Chang:2017cdx,Chang:2017mxc}, central charges of five-dimensional superconformal field theories were shown to be related to deformations of the (round) five-sphere partition function. In particular, the conformal central charge $C_T$, defined (in $d$-dimensions) as the coefficient in an appropriately normalized stress-tensor two-point function
\ie
\left\langle T_{\mu\nu} (x) T_{\rho\sigma} (0)\right\rangle \ = \ \frac{C_{T}}{V_{{\rm \hat S}^{d-1}}^{2}}\frac{\cI_{\mu\nu,\rho\sigma}(x)}{x^{2d}} \,,
\fe
where $V_{\hat {\rm S}^{d-1}} = \frac{2 \pi^{d/2}}{\Gamma(d/2)}$ is the unit volume of a $(d-1)$-dimensional sphere, and $\cI_{\mu\nu,\rho\sigma}(x)$ is the conformally covariant structure
\ie
\cI_{\mu\nu,\rho\sigma}(x) \ = \ & \frac{1}{2} \left[ I_{\mu\sigma} (x) I_{\nu\rho}(x) + I_{\mu \rho}(x) I_{\nu \sigma} (x) \right] - \frac{1}{d}\delta_{\mu\nu}\delta_{\rho\sigma} \,,\\
I_{\mu\nu}(x) \ = \ & \delta_{\mu\nu} - 2 \frac{x_{\mu}x_{\nu}}{x^{2}} \,,
\fe
can be computed from the squashed five-sphere partition function. More precisely, the conformal central charge $C_T$ is related to the squashed five-sphere free energy 
\ie
 F \left[\omega_1,\omega_2,\omega_3\right] \ = \ - \log \cZ \left[\omega_1,\omega_2,\omega_3\right] 
\fe
as follows
\ie\label{eqn:CTvsF}
 F \left[(1+a_1),(1+a_2),(1+a_3)\right] \ = \ F_{\rm S^{5}} - \frac{\pi^{2} C_T}{1920} \left( \sum_{i=1}^{3} a_i^{2} - \sum_{i<j} a_i a_j \right) + \cO(a^{3}) \,,
\fe
and thus can be computed by expanding the latter to second order in the squashing parameters $a_i$. We employ this relation to find explicit numerical results for $C_T$ which we compare to supergravity predictions in the main text.

\section{Numerical results for \texorpdfstring{$\#_{N,M}$}{D5/NS5} theories}
\label{App:Numerics}

In this section we present the values for the round five-sphere partition function $-F_{\rm S^{5}}^{\#_{N,M}}$ as well as the conformal central charges $C_T^{\#_{N,M}}$ for the quiver deformations of the $\#_{N,M}$ theories. They are computed using numerical saddle point methods, as explained in Section~\ref{sec:largeN}.

\begin{table}[H]
\centering
{\renewcommand{\arraystretch}{1.15}
\begin{tabular}{|c||cccccccc|}
\hline
$ -F_{\rm S^5}^{\#_{N,M}}  $ & $ ({M,2}) $ & $ ({M,4}) $ & $ ({M,6}) $ & $ ({M,8}) $ & $ ({M,10}) $ & $ ({M,12}) $ & $ ({M,14}) $ & $ ({M,16}) $ \\
\hline
\hline
 $({2,N}) $ & $ 11.0521 $ & $ 55.9644 $ & $ 132.180 $ & $ 239.646 $ & $ 378.340 $ & $ 548.251 $ & $ 749.373 $ & $ 981.702 $ \\
 $({4,N}) $ & $ 55.5040 $ & $ 292.953 $ & $ 696.826 $ & $ 1266.13 $ & $ 2000.72 $ & $ 2900.52 $ & $ 3965.52 $ & $ 5195.70 $ \\
 $({6,N}) $ & $ 129.005 $ & $ 690.652 $ & $ 1652.01 $ & $ 3008.18 $ & $ 4757.98 $ & $ 6901.04 $ & $ 9437.21 $ & $ 12366.4 $ \\
 $({8,N}) $ & $ 230.754 $ & $ 1244.40 $ & $ 2988.21 $ & $ 5451.60 $ & $ 8631.05 $ & $ 12525.2 $ & $ 17133.5 $ & $ 22455.7 $ \\
 $({10,N}) $ & $ 360.400 $ & $ 1952.13 $ & $ 4700.50 $ & $ 8588.59 $ & $ 13609.5 $ & $ 19760.1 $ & $ 27039.1 $ & $ 35445.7 $ \\
 $({12,N}) $ & $ 517.746 $ & $ 2812.64 $ & $ 6785.93 $ & $ 12414.1 $ & $ 19686.1 $ & $ 28596.8 $ & $ 39143.2 $ & $ 51324.0 $ \\
 $({14,N}) $ & $ 702.663 $ & $ 3825.19 $ & $ 9242.54 $ & $ 16924.5 $ & $ 26855.9 $ & $ 39028.4 $ & $ 53437.7 $ & $ 70081.2 $ \\
 $({16,N}) $ & $ 915.064 $ & $ 4989.26 $ & $ 12069.0 $ & $ 22117.4 $ & $ 35114.8 $ & $ 51049.9 $ & $ 69916.2 $ & $ 91709.5 $ \\
 $({18,N}) $ & $ 1154.88 $ & $ 6304.48 $ & $ 15264.2 $ & $ 27990.9 $ & $ 44460.1 $ & $ 64657.1 $ & $ 88573.2 $ & $ 116203. $ \\
 $({20,N}) $ & $ 1422.07 $ & $ 7770.54 $ & $ 18827.5 $ & $ 34543.5 $ & $ 54889.2 $ & $ 79846.7 $ & $ 109405. $ & $ 143556. $ \\
 $({22,N}) $ & $ 1716.59 $ & $ 9387.24 $ & $ 22758.3 $ & $ 41774.0 $ & $ 66400.4 $ & $ 96616.1 $ & $ 132407. $ & $ 173764. $ \\
 $({24,N}) $ & $ 2038.41 $ & $ 11154.4 $ & $ 27056.0 $ & $ 49681.5 $ & $ 78992.2 $ & $ 114963. $ & $ 157577. $ & $ 206823. $ \\
 $({26,N}) $ & $ 2387.51 $ & $ 13071.8 $ & $ 31720.3 $ & $58265.2 $ & $ 92663.1 $ & $ 134885. $ & $ 184913. $ & $ 242731. $ \\
 $({28,N}) $ & $ 2763.85 $ & $ 15139.4 $ & $ 36750.7 $ & $67524.4 $ & $ 107412. $ & $ 156382. $ & $ 214411. $ & $ 281484. $ \\
 $({30,N}) $ & $ 3167.42 $ & $ 17357.1 $ & $ 42147.2 $ & $ 77458.6 $ & $ 123239. $ & $ 179451. $ & $ 246071. $ & $ 323080. $ \\
  \hline
 \end{tabular}
\caption{Explicit numerical values for the saddle point evaluation of the five-sphere free energy $-F_{\rm S^5}^{\#_{N,M}} $ for five-dimensional $\#_{N,M}$ theories for even $2 \leq M \leq 30$ and even $2 \leq N \leq 16$. For odd values of $M,N$, the data behaves similarly and are depicted in the corresponding plots given in Figures~\ref{fig:hash-fit} and~\ref{Fig:LogLogFvsSugraM}.}\label{tbl:HTF1}}
 \end{table}
 
 \begin{table}[H]
\centering
{\renewcommand{\arraystretch}{1.15}
\begin{tabular}{|c||ccccccc|}
\hline
$ -F_{\rm S^5}^{\#_{N,M}}$ & $ ({M,18}) $ & $({M,20})$ & $({M,22})$ & $({M,24})$ & $({M,26})$ & $({M,28})$ & $({M,30})$ \\
\hline
\hline
 $ ({2,N}) $ & $ 1245.24 $ & $ 1539.97 $ & $ 1865.91 $ & $ 2223.05 $ & $ 2611.39 $ & $ 3030.94 $ & $ 3481.68 $ \\
 $ ({4,N}) $ & $ 6591.04 $ & $ 8151.54 $ & $ 9877.20 $ & $ 11768.0 $ & $ 13824.0 $ & $ 16045.1 $ & $ 18431.3 $ \\
 $ ({6,N}) $ & $ 15688.6 $ & $ 19403.7 $ & $ 23511.8 $ & $ 28012.8 $ & $ 32906.8 $ & $ 38193.6 $ & $ 43873.4 $ \\
 $ ({8,N}) $ & $ 28491.7 $ & $ 35241.2 $ & $ 42704.3 $ & $ 50881.0 $ & $ 59771.2 $ & $ 69374.9 $ & $ 79692.1 $ \\
 $ ({10,N}) $ & $ 44979.5 $ & $ 55640.2 $ & $ 67427.8 $ & $ 80342.1 $ & $ 94383.0 $ & $ 109550. $ & $ 125845. $ \\
 $ ({12,N}) $ & $ 65138.2 $ & $ 80585.4 $ & $ 97665.1 $ & $ 116377. $ & $ 136721. $ & $ 158698. $ & $ 182306. $ \\
 $ ({14,N}) $ & $ 88957.1 $ & $ 110065. $ & $ 133403. $ & $ 158972. $ & $ 186771. $ & $ 216800. $ & $ 249060. $ \\
 $ ({16,N}) $ & $ 116428. $ & $ 144068. $ & $ 174631. $ & $ 208115. $ & $ 244520. $ & $ 283845. $ & $ 326090. $ \\
 $ ({18,N}) $ & $ 147542. $ & $ 182588. $ & $ 221341. $ & $ 263797. $ & $ 309957. $ & $ 359821. $ & $ 413387. $ \\
 $ ({20,N}) $ & $ 182294. $ & $ 225618. $ & $ 273523. $ & $ 326009. $ & $ 383074. $ & $ 444717. $ & $ 510938. $ \\
 $ ({22,N}) $ & $ 220679. $ & $ 273150. $ & $ 331172. $ & $ 394743. $ & $ 463861. $ & $ 538525. $ & $ 618735. $ \\
 $ ({24,N}) $ & $ 262693. $ & $ 325180. $ & $ 394281. $ & $ 469992. $ & $ 552312. $ & $ 641238. $ & $ 736769. $ \\
 $ ({26,N}) $ & $ 308330. $ & $ 381703. $ & $ 462845. $ & $ 551751. $ & $ 648419. $ & $ 752847. $ & $ 865032. $ \\
 $ ({28,N}) $ & $ 357588. $ & $ 442716. $ & $ 536859. $ & $ 640015. $ & $ 752178. $ & $ 873346. $ & $ 1.00352\cdot10^6 $ \\
 $ ({30,N}) $ & $ 410464. $ & $ 508213. $ & $ 616320. $ & $ 734778. $ & $ 863582. $ & $ 1.00273\cdot10^6 $ & $ 1.15222\cdot10^6$\\
  \hline
 \end{tabular}
 \caption{Explicit numerical values for the saddle point evaluation of the five-sphere free energy $-F_{\rm S^5}^{\#_{N,M}} $ for five-dimensional $\#_{N,M}$ theories for even $2 \leq M \leq 30$ and even $18 \leq N \leq 30$. For odd values of $M,N$, the data behaves similarly and are depicted in the corresponding plots given in Figures~\ref{fig:hash-fit} and~\ref{Fig:LogLogFvsSugraM}.} \label{tbl:HTF2}}
 \end{table}

\begin{table}[H]
\centering
{\renewcommand{\arraystretch}{1.1}
\setlength{\tabcolsep}{5pt}
{\small
\begin{tabular}{|c||cccccccc|}
\hline
$ C_{T}^{\#_{N,M}}  $ & $ ({M,2}) $ & $ ({M,3}) $ & $ ({M,4}) $ & $ ({M,5}) $ & $ ({M,6}) $ & $ ({M,7}) $ & $ ({M,8}) $ & $ ({M,9}) $ \\
\hline
\hline
$({2,N}) $ & $ 867.477 $ & $ 2182.92 $ & $ 3978.62 $ & $ 6242.7 $ & $ 8968.39 $ & $ 12151.3 $ & $ 15788.4 $ & $ 19877.4 $ \\
 $({3,N}) $ & $ 2242.68 $ & $ 5803.35 $ & $ 10730.3 $ & $ 16985.8 $ & $ 24549.0 $ & $ 33406.7 $ & $ 43549.9 $ & $ 54971.9 $ \\
 $({4,N}) $ & $ 4163.03 $ & $ 10893.6 $ & $ 20289.7 $ & $ 32280.1$ & $ 46823.8 $ & $ 63894.4 $ & $ 83473.3 $ & $ 105547. $ \\
 $({5,N}) $ & $ 6628.07 $ & $ 17440.8 $ & $ 32615.3 $ & $ 52046.2 $ & $ 75671.5 $ & $ 103449. $ & $ 135350. $ & $ 171350. $ \\
 $({6,N}) $ & $ 9638.98 $ & $ 25441.6 $ & $ 47691.3 $ & $ 76250.4 $ & $ 111035. $ & $ 151987. $ & $ 199066. $ & $ 252238. $ \\
 $({7,N}) $ & $ 13196.5 $ & $ 34895.8 $ & $ 65512.2 $ & $ 104876.$ & $ 152883. $ & $ 209461. $ & $ 274554. $ & $ 348120. $ \\
 $({8,N}) $ & $ 17300.9 $ & $ 45803.7 $ & $ 86075.5 $ & $ 137916. $ & $ 201201. $ & $ 275842. $ & $ 361772. $ & $ 458938. $ \\
 $({9,N}) $ & $ 21952.4 $ & $ 58165.5 $ & $ 109381. $ & $ 175366.$ & $ 255979. $ & $ 351115. $ & $ 460695. $ & $ 584657. $ \\
 $({10,N}) $ & $ 27151.0 $ & $ 71981.7 $ & $ 135428. $ & $ 217225. $ & $ 317211. $ & $ 435268. $ & $ 571305. $ & $ 725251. $ \\
 $({11,N}) $ & $ 32896.7 $ & $ 87252.2 $ & $ 164217. $ & $263492. $ & $ 384895. $ & $ 528296. $ & $ 693593. $ & $ 880703. $ \\
 $({12,N}) $ & $ 39189.6 $ & $ 103977. $ & $ 195748. $ & $314166. $ & $ 459030. $ & $ 630195. $ & $ 827549. $ & $ 1.05100\cdot10^6 $ \\
 $({13,N}) $ & $ 46029.5 $ & $ 122156. $ & $ 230021. $ & $369248. $ & $ 539613. $ & $ 740961. $ & $ 973170. $ & $ 1.23614\cdot10^6 $ \\
 $({14,N}) $ & $ 53408.2 $ & $ 141789. $ & $ 267037. $ & $428736. $ & $ 626645. $ & $ 860595. $ & $ 1.13045\cdot10^6 $ & $ 1.43611\cdot10^6 $ \\
 $({15,N}) $ & $ 61351.3 $ & $ 162877. $ & $ 306794. $ & $ 492632. $ & $ 720126. $ & $ 989094. $ & $ 1.29939\cdot10^6 $ & $1.65091\cdot10^6 $ \\
  \hline
 \end{tabular}}
\caption{Explicit numerical values for the saddle point evaluation of the conformal central charge $C_T^{\#_{N,M}}$ for five-dimensional $\#_{N,M}$ theories for $2 \leq M \leq 15$ and $2 \leq N \leq 9$.}\label{tbl:HTCT1}}
 \end{table}

\begin{table}[H]
\centering
{\renewcommand{\arraystretch}{1.15}
\setlength{\tabcolsep}{5pt}
{\small
\begin{tabular}{|c||cccccc|}
\hline
$ C_{T}^{\#_{N,M}}  $ & $ ({M,10}) $ & $ ({M,11}) $ & $ ({M,12}) $ & $ ({M,13}) $ & $ ({M,14}) $ & $ ({M,15}) $ \\
\hline
\hline
$({2,N}) $ & $ 24416.6 $ & $ 29404.7 $ & $ 34840.6 $ & $ 40723.3$ & $ 47052.2 $ & $ 53826.5 $ \\
$({3,N}) $ & $ 67667.8 $ & $ 81633.6 $ & $ 96866.3 $ & $ 113363.$ & $ 131122. $ & $ 150141. $ \\
$({4,N}) $ & $ 130105. $ & $ 157140. $ & $ 186644. $ & $ 218613.$ & $ 253041. $ & $ 289926. $ \\
$({5,N}) $ & $ 211434. $ & $ 255586. $ & $ 303796. $ & $ 356054.$ & $ 412353. $ & $ 472687. $ \\
$({6,N}) $ & $ 311478. $ & $ 376765. $ & $ 448083. $ & $ 525418.$ & $ 608757. $ & $ 698091. $ \\
$({7,N}) $ & $ 430124. $ & $ 520540. $ & $ 619343. $ & $ 726514.$ & $ 842037. $ & $ 965898. $ \\
$({8,N}) $ & $ 567298. $ & $ 686816. $ & $ 817461. $ & $ 959209.$ & $ 1.11204\cdot10^6 $ & $ 1.27593\cdot10^6 $ \\
$({9,N}) $ & $ 722949. $ & $ 875529. $ & $ 1.04236\cdot10^6 $ &$ 1.22341\cdot10^6 $ & $ 1.41865\cdot10^6 $ & $ 1.62806\cdot10^6 $\\
$({10,N}) $ & $ 897044. $ & $ 1.08664\cdot10^6 $ & $ 1.29398\cdot10^6 $ & $ 1.51904\cdot10^6 $ & $ 1.76179\cdot10^6 $ & $2.02219\cdot10^6 $ \\
$({11,N}) $ & $ 1.08956\cdot10^6 $ & $ 1.3201\cdot10^6 $ & $1.57228\cdot10^6 $ & $ 1.84606\cdot10^6 $ & $ 2.14139\cdot10^6 $ &$ 2.45825\cdot10^6 $ \\
$({12,N}) $ & $ 1.30048\cdot10^6 $ & $ 1.57591\cdot10^6 $ & $1.87724\cdot10^6 $ & $ 2.20442\cdot10^6 $ & $ 2.55741\cdot10^6 $ &$ 2.93618\cdot10^6 $ \\
$({13,N}) $ & $ 1.52978\cdot10^6 $ & $ 1.85403\cdot10^6 $ & $2.20883\cdot10^6 $ & $ 2.5941\cdot10^6 $ & $ 3.00982\cdot10^6 $ & $ 3.45593\cdot10^6 $ \\
$({14,N}) $ & $ 1.77747\cdot10^6 $ & $ 2.15447\cdot10^6 $ & $2.56702\cdot10^6 $ & $ 3.01508\cdot10^6 $ & $ 3.49858\cdot10^6 $ &$ 4.01748\cdot10^6 $ \\
$({15,N}) $ & $ 2.04354\cdot10^6 $ & $ 2.4772\cdot10^6 $ & $2.95182\cdot10^6 $ & $ 3.46733\cdot10^6 $ & $ 4.02368\cdot10^6 $ &$ 4.6208\times10^6 $ \\
  \hline
 \end{tabular}}
\caption{Explicit numerical values for the saddle point evaluation of the conformal central charge $C_T^{\#_{N,M}}$ for five-dimensional $\#_{N,M}$ theories for $2 \leq M \leq 15$ and $10 \leq N \leq 15$.}\label{tbl:HTCT2}}
 \end{table}

 \begin{figure}[H]
 \centering
  \includegraphics[width=0.95\linewidth]{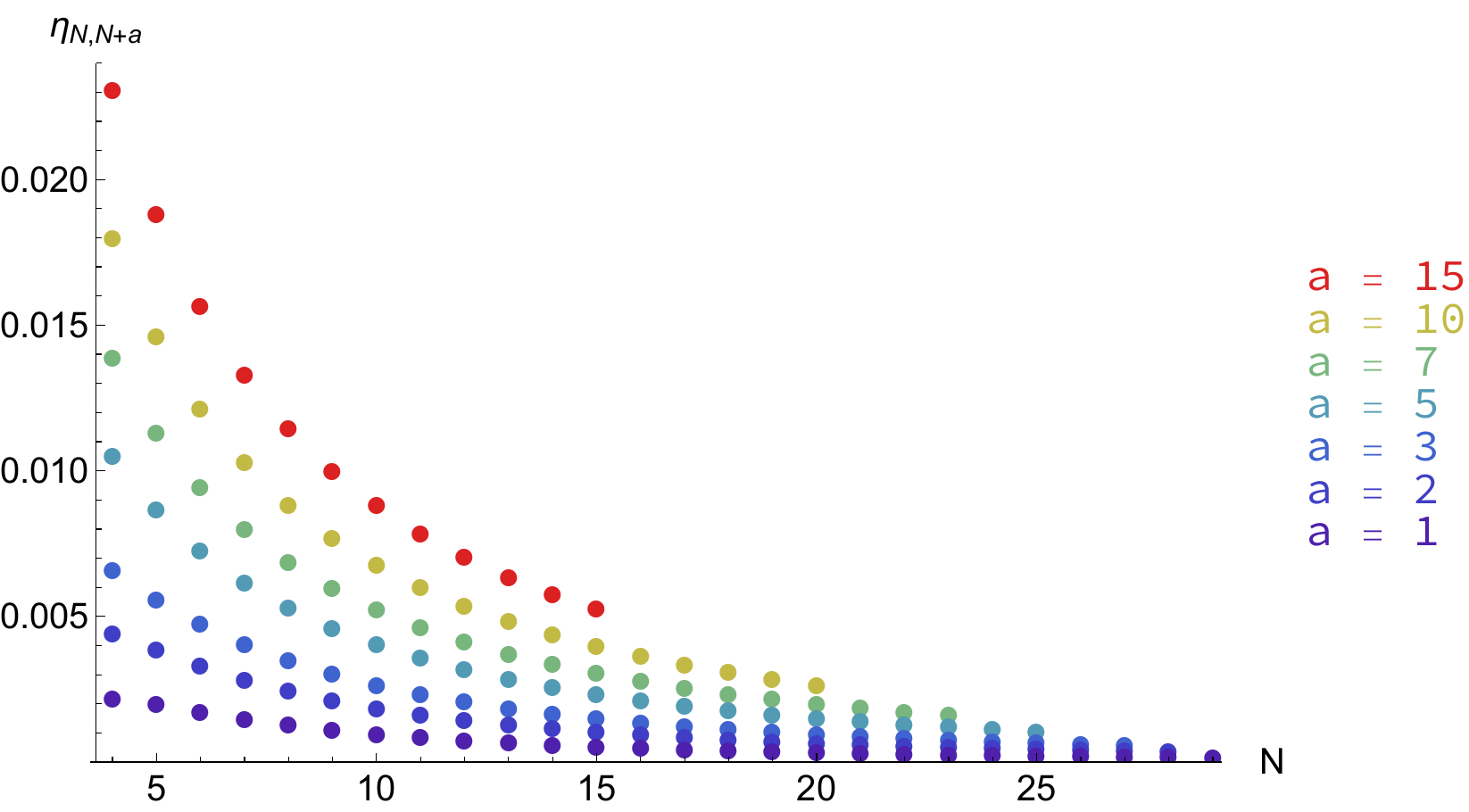}
 \caption{The relative differences between the free energies of S-dual $\#_{N,N+a}$ superconformal field theories, with $\eta_{M,N}$ as defined in equation~\eqref{eqn:etadef}. As expected and argued in the main text, the relative difference is small (within $\cO(3\%)$ even for small values of $N$ and $a=15$) and shows converging behavior for increasing $N$. Notice that since the data presented in this paper is restricted to $M,N \leq 30$, we are only able to show $\eta_{N,N+a}$ up to $N=30-a$.}
 \label{Fig:Sduality}
\end{figure}

\section{\texorpdfstring{$C_T/F_{S^5}$}{C-T/F} in Kaluza-Klein supergravity}\label{sec:app-CT}

The relation between the conformal central charge $C_T$ and the five-sphere partition function discussed in Section~\ref{sec:CT-IIB} holds for generic holographic conformal field theories in odd dimensions $d$ whose dual admits a (consistent) Kaluza-Klein reduction on ${\rm AdS}_{d+1}$ of the form 
\ie
 S_{d+1}& \ = \ \frac{1}{2\kappa_{d+1}^2}\int_{{\rm AdS}_{d+1}} \diff^{d+1}x \sqrt{-g}\left[R_{d+1}-2\Lambda_{d+1}+\dots\right]\,,
\fe
with effective $(d+1)$-dimensional gravitational coupling $\kappa^2_{d+1}$ and the dots denoting matter fields which vanish in the ${\rm AdS}_{d+1}$ vacuum. In particular, in the ${\rm AdS}_{d+1}$ vacuum there are the following relations
\begin{align}
 R_{d+1}& \ = \ \frac{2(d+1)}{d-1}\Lambda_{d+1}\,, & \Lambda_{d+1}& \ = \ -\frac{d(d-1)}{2L}\,,
\end{align}
with $L$ the AdS$_{d+1}$ curvature radius.
With ${\rm AdS}_{d+1}$ in global coordinates, 
\ie
\diff s^2 \ = \ \diff u^2+L^2\sinh^2\!u\,\diff s^2_{{\rm S}^d}\,,
\fe
the $d$-dimensional (unit) sphere partition function is obtained from this reduced action as
\ie
 F_{{\rm S}^d}& \ = \ -S_{d+1}\big\vert_{\rm AdS_{d+1}} \ = \ \frac{d L^{d-1}}{\kappa_{d+1}^2} \Vol_{{\rm AdS}_{d+1}}\,,
\fe
where $\Vol_{{\rm AdS}_{d+1}}$ denotes the renormalized volume of unit radius AdS$_{d+1}$. 
For odd dimensional spheres ${\rm S}^d$, this quantity is well defined and given by the finite part of the integral 
\ie
V_{{\rm S}^d}\int_0^{\arcsin(1/\epsilon)} \diff u \, \sinh^du \,, \quad \text{for} \quad (\epsilon\searrow 0) \,.
\fe
With the formula for $C_T$ as given in~\cite{Freedman:1998tz,Penedones:2016voo} (see also Section~4.1 of~\cite{Chang:2017mxc}), we find
\ie
 \frac{C_T}{F_{\rm S^5}}& \ = \ \frac{4\pi^{d/2}\Gamma(d+2)}{d(d-1)\Gamma\big(\frac{d}{2}\big)^3\Vol_{{\rm AdS}_{d+1}}} \ = \ \begin{cases}
                                                                                                 ~\displaystyle{-\frac{640}{\pi^{2}}} \,, & d=5 \,, \\[4mm]
                                                                                                 ~\displaystyle{+\frac{96}{\pi^{2}}} \,, & d=3 \,.
                                                                                                \end{cases}
\fe
This is a universal relation for holographic conformal field theories whose duals are of this generic form, and is analogous to the $a=c$ relation between the central charges for holographic conformal field theories in four dimensions. 

\bibliographystyle{JHEP}
\bibliography{ads6}

\providecommand{\href}[2]{#2}\begingroup\raggedright\begin{thebibliography}{10}

\bibitem{Seiberg:1996bd}
N.~Seiberg, \emph{{Five-dimensional SUSY field theories, nontrivial fixed
  points and string dynamics}},
  \href{http://dx.doi.org/10.1016/S0370-2693(96)01215-4}{\emph{Phys. Lett.}
  {\bf B388} (1996) 753--760}, [\href{http://arxiv.org/abs/hep-th/9608111}{{\tt
  hep-th/9608111}}].

\bibitem{Morrison:1996xf}
D.~R. Morrison and N.~Seiberg, \emph{{Extremal transitions and five-dimensional
  supersymmetric field theories}},
  \href{http://dx.doi.org/10.1016/S0550-3213(96)00592-5}{\emph{Nucl. Phys.}
  {\bf B483} (1997) 229--247}, [\href{http://arxiv.org/abs/hep-th/9609070}{{\tt
  hep-th/9609070}}].

\bibitem{Intriligator:1997pq}
K.~A. Intriligator, D.~R. Morrison and N.~Seiberg, \emph{{Five-dimensional
  supersymmetric gauge theories and degenerations of Calabi-Yau spaces}},
  \href{http://dx.doi.org/10.1016/S0550-3213(97)00279-4}{\emph{Nucl. Phys.}
  {\bf B497} (1997) 56--100}, [\href{http://arxiv.org/abs/hep-th/9702198}{{\tt
  hep-th/9702198}}].

\bibitem{Aharony:1997ju}
O.~Aharony and A.~Hanany, \emph{{Branes, superpotentials and superconformal
  fixed points}},
  \href{http://dx.doi.org/10.1016/S0550-3213(97)00472-0}{\emph{Nucl. Phys.}
  {\bf B504} (1997) 239--271}, [\href{http://arxiv.org/abs/hep-th/9704170}{{\tt
  hep-th/9704170}}].

\bibitem{Aharony:1997bh}
O.~Aharony, A.~Hanany and B.~Kol, \emph{{Webs of (p,q) five-branes,
  five-dimensional field theories and grid diagrams}},
  \href{http://dx.doi.org/10.1088/1126-6708/1998/01/002}{\emph{JHEP} {\bf 01}
  (1998) 002}, [\href{http://arxiv.org/abs/hep-th/9710116}{{\tt
  hep-th/9710116}}].

\bibitem{Brandhuber:1999np}
A.~Brandhuber and Y.~Oz, \emph{{The D-4 - D-8 brane system and five-dimensional
  fixed points}},
  \href{http://dx.doi.org/10.1016/S0370-2693(99)00763-7}{\emph{Phys. Lett.}
  {\bf B460} (1999) 307--312}, [\href{http://arxiv.org/abs/hep-th/9905148}{{\tt
  hep-th/9905148}}].

\bibitem{Ferrara:1998gv}
S.~Ferrara, A.~Kehagias, H.~Partouche and A.~Zaffaroni, \emph{{AdS(6)
  interpretation of 5-D superconformal field theories}},
  \href{http://dx.doi.org/10.1016/S0370-2693(98)00560-7}{\emph{Phys. Lett.}
  {\bf B431} (1998) 57--62}, [\href{http://arxiv.org/abs/hep-th/9804006}{{\tt
  hep-th/9804006}}].

\bibitem{Bergman:2012kr}
O.~Bergman and D.~Rodriguez-Gomez, \emph{{5d quivers and their AdS(6) duals}},
  \href{http://dx.doi.org/10.1007/JHEP07(2012)171}{\emph{JHEP} {\bf 1207}
  (2012) 171}, [\href{http://arxiv.org/abs/1206.3503}{{\tt 1206.3503}}].

\bibitem{Jafferis:2012iv}
D.~L. Jafferis and S.~S. Pufu, \emph{{Exact results for five-dimensional
  superconformal field theories with gravity duals}},
  \href{http://dx.doi.org/10.1007/JHEP05(2014)032}{\emph{JHEP} {\bf 05} (2014)
  032}, [\href{http://arxiv.org/abs/1207.4359}{{\tt 1207.4359}}].

\bibitem{Bergman:2012qh}
O.~Bergman and D.~Rodriguez-Gomez, \emph{{Probing the Higgs branch of 5d fixed
  point theories with dual giant gravitons in AdS(6)}},
  \href{http://dx.doi.org/10.1007/JHEP12(2012)047}{\emph{JHEP} {\bf 12} (2012)
  047}, [\href{http://arxiv.org/abs/1210.0589}{{\tt 1210.0589}}].

\bibitem{Passias:2012vp}
A.~Passias, \emph{{A note on supersymmetric AdS$_6$ solutions of massive type
  IIA supergravity}},
  \href{http://dx.doi.org/10.1007/JHEP01(2013)113}{\emph{JHEP} {\bf 01} (2013)
  113}, [\href{http://arxiv.org/abs/1209.3267}{{\tt 1209.3267}}].

\bibitem{Assel:2012nf}
B.~Assel, J.~Estes and M.~Yamazaki, \emph{{Wilson Loops in 5d N=1 SCFTs and
  AdS/CFT}}, \href{http://dx.doi.org/10.1007/s00023-013-0249-5}{\emph{Annales
  Henri Poincare} {\bf 15} (2014) 589--632},
  [\href{http://arxiv.org/abs/1212.1202}{{\tt 1212.1202}}].

\bibitem{Bergman:2013koa}
O.~Bergman, D.~Rodr{\'\i}guez-G{\'o}mez and G.~Zafrir, \emph{{5d superconformal
  indices at large N and holography}},
  \href{http://dx.doi.org/10.1007/JHEP08(2013)081}{\emph{JHEP} {\bf 08} (2013)
  081}, [\href{http://arxiv.org/abs/1305.6870}{{\tt 1305.6870}}].

\bibitem{Alday:2014rxa}
L.~F. Alday, M.~Fluder, P.~Richmond and J.~Sparks, \emph{{Gravity Dual of
  Supersymmetric Gauge Theories on a Squashed Five-Sphere}},
  \href{http://dx.doi.org/10.1103/PhysRevLett.113.141601}{\emph{Phys. Rev.
  Lett.} {\bf 113} (2014) 141601}, [\href{http://arxiv.org/abs/1404.1925}{{\tt
  1404.1925}}].

\bibitem{Alday:2014bta}
L.~F. Alday, M.~Fluder, C.~M. Gregory, P.~Richmond and J.~Sparks,
  \emph{{Supersymmetric gauge theories on squashed five-spheres and their
  gravity duals}}, \href{http://dx.doi.org/10.1007/JHEP09(2014)067}{\emph{JHEP}
  {\bf 09} (2014) 067}, [\href{http://arxiv.org/abs/1405.7194}{{\tt
  1405.7194}}].

\bibitem{Alday:2014fsa}
L.~F. Alday, P.~Richmond and J.~Sparks, \emph{{The holographic supersymmetric
  Renyi entropy in five dimensions}},
  \href{http://dx.doi.org/10.1007/JHEP02(2015)102}{\emph{JHEP} {\bf 02} (2015)
  102}, [\href{http://arxiv.org/abs/1410.0899}{{\tt 1410.0899}}].

\bibitem{Hama:2014iea}
N.~Hama, T.~Nishioka and T.~Ugajin, \emph{{Supersymmetric R{\'e}nyi entropy in
  five dimensions}},
  \href{http://dx.doi.org/10.1007/JHEP12(2014)048}{\emph{JHEP} {\bf 12} (2014)
  048}, [\href{http://arxiv.org/abs/1410.2206}{{\tt 1410.2206}}].

\bibitem{Alday:2015jsa}
L.~F. Alday, M.~Fluder, C.~M. Gregory, P.~Richmond and J.~Sparks,
  \emph{{Supersymmetric solutions to Euclidean Romans supergravity}},
  \href{http://dx.doi.org/10.1007/JHEP02(2016)100}{\emph{JHEP} {\bf 02} (2016)
  100}, [\href{http://arxiv.org/abs/1505.04641}{{\tt 1505.04641}}].

\bibitem{DHoker:2016ujz}
E.~D'Hoker, M.~Gutperle, A.~Karch and C.~F. Uhlemann, \emph{{Warped
  $AdS_6\times S^2$ in Type IIB supergravity I: Local solutions}},
  \href{http://dx.doi.org/10.1007/JHEP08(2016)046}{\emph{JHEP} {\bf 08} (2016)
  046}, [\href{http://arxiv.org/abs/1606.01254}{{\tt 1606.01254}}].

\bibitem{DHoker:2016ysh}
E.~D'Hoker, M.~Gutperle and C.~F. Uhlemann, \emph{{Holographic duals for
  five-dimensional superconformal quantum field theories}},
  \href{http://dx.doi.org/10.1103/PhysRevLett.118.101601}{\emph{Phys. Rev.
  Lett.} {\bf 118} (2017) 101601}, [\href{http://arxiv.org/abs/1611.09411}{{\tt
  1611.09411}}].

\bibitem{DHoker:2017mds}
E.~D'Hoker, M.~Gutperle and C.~F. Uhlemann, \emph{{Warped $AdS_6\times S^2$ in
  Type IIB supergravity II: Global solutions and five-brane webs}},
  \href{http://dx.doi.org/10.1007/JHEP05(2017)131}{\emph{JHEP} {\bf 05} (2017)
  131}, [\href{http://arxiv.org/abs/1703.08186}{{\tt 1703.08186}}].

\bibitem{DHoker:2017zwj}
E.~D'Hoker, M.~Gutperle and C.~F. Uhlemann, \emph{{Warped $AdS_6\times S^2$ in
  Type IIB supergravity III: Global solutions with seven-branes}},
  \href{http://dx.doi.org/10.1007/JHEP11(2017)200}{\emph{JHEP} {\bf 11} (2017)
  200}, [\href{http://arxiv.org/abs/1706.00433}{{\tt 1706.00433}}].

\bibitem{Gutperle:2017tjo}
M.~Gutperle, C.~Marasinou, A.~Trivella and C.~F. Uhlemann, \emph{{Entanglement
  entropy vs. free energy in IIB supergravity duals for 5d SCFTs}},
  \href{http://dx.doi.org/10.1007/JHEP09(2017)125}{\emph{JHEP} {\bf 09} (2017)
  125}, [\href{http://arxiv.org/abs/1705.01561}{{\tt 1705.01561}}].

\bibitem{Gutperle:2018vdd}
M.~Gutperle, A.~Trivella and C.~F. Uhlemann, \emph{{Type IIB 7-branes in warped
  $AdS_6$: partition functions, brane webs and probe limit}},
  \href{http://arxiv.org/abs/1802.07274}{{\tt 1802.07274}}.

\bibitem{Gutperle:2018wuk}
M.~Gutperle, C.~F. Uhlemann and O.~Varela, \emph{{Massive Spin 2 excitations in
  $AdS_6 \times S^2$ warped spacetimes}},
  \href{http://arxiv.org/abs/1805.11914}{{\tt 1805.11914}}.

\bibitem{Kaidi:2017bmd}
J.~Kaidi, \emph{{(p,q)-strings probing five-brane webs}},
  \href{http://dx.doi.org/10.1007/JHEP10(2017)087}{\emph{JHEP} {\bf 10} (2017)
  087}, [\href{http://arxiv.org/abs/1708.03404}{{\tt 1708.03404}}].

\bibitem{Bergman:2018}
O.~Bergman, D.~Rodr{\'\i}guez-G{\'o}mez and C.~F. Uhlemann, \emph{{Testing
  $AdS_6/CFT_5$ in Type IIB with stringy operators}},
  \href{http://arxiv.org/abs/1806.XXXXX}{{\tt 1806.XXXXX}}.

\bibitem{Pestun:2007rz}
V.~Pestun, \emph{{Localization of gauge theory on a four-sphere and
  supersymmetric Wilson loops}},
  \href{http://dx.doi.org/10.1007/s00220-012-1485-0}{\emph{Commun. Math. Phys.}
  {\bf 313} (2012) 71--129}, [\href{http://arxiv.org/abs/0712.2824}{{\tt
  0712.2824}}].

\bibitem{Kallen:2012va}
J.~K{\"a}ll{\'e}n, J.~Qiu and M.~Zabzine, \emph{{The perturbative partition
  function of supersymmetric 5D Yang-Mills theory with matter on the
  five-sphere}}, \href{http://dx.doi.org/10.1007/JHEP08(2012)157}{\emph{JHEP}
  {\bf 08} (2012) 157}, [\href{http://arxiv.org/abs/1206.6008}{{\tt
  1206.6008}}].

\bibitem{Kim:2012ava}
H.-C. Kim and S.~Kim, \emph{{M5-branes from gauge theories on the 5-sphere}},
  \href{http://dx.doi.org/10.1007/JHEP05(2013)144}{\emph{JHEP} {\bf 05} (2013)
  144}, [\href{http://arxiv.org/abs/1206.6339}{{\tt 1206.6339}}].

\bibitem{Imamura:2012bm}
Y.~Imamura, \emph{{Perturbative partition function for squashed $S^5$}},
  \href{http://dx.doi.org/10.1093/ptep/ptt044}{\emph{PTEP} {\bf 2013} (2013)
  073B01}, [\href{http://arxiv.org/abs/1210.6308}{{\tt 1210.6308}}].

\bibitem{Lockhart:2012vp}
G.~Lockhart and C.~Vafa, \emph{{Superconformal Partition Functions and
  Non-perturbative Topological Strings}},
  \href{http://arxiv.org/abs/1210.5909}{{\tt 1210.5909}}.

\bibitem{Imamura:2012xg}
Y.~Imamura, \emph{{Supersymmetric theories on squashed five-sphere}},
  \href{http://dx.doi.org/10.1093/ptep/pts052}{\emph{PTEP} {\bf 2013} (2013)
  013B04}, [\href{http://arxiv.org/abs/1209.0561}{{\tt 1209.0561}}].

\bibitem{Benini:2009gi}
F.~Benini, S.~Benvenuti and Y.~Tachikawa, \emph{{Webs of five-branes and N=2
  superconformal field theories}},
  \href{http://dx.doi.org/10.1088/1126-6708/2009/09/052}{\emph{JHEP} {\bf 09}
  (2009) 052}, [\href{http://arxiv.org/abs/0906.0359}{{\tt 0906.0359}}].

\bibitem{Bao:2013pwa}
L.~Bao, V.~Mitev, E.~Pomoni, M.~Taki and F.~Yagi, \emph{{Non-Lagrangian
  Theories from Brane Junctions}},
  \href{http://dx.doi.org/10.1007/JHEP01(2014)175}{\emph{JHEP} {\bf 01} (2014)
  175}, [\href{http://arxiv.org/abs/1310.3841}{{\tt 1310.3841}}].

\bibitem{Gaiotto:2009we}
D.~Gaiotto, \emph{{N=2 dualities}},
  \href{http://dx.doi.org/10.1007/JHEP08(2012)034}{\emph{JHEP} {\bf 08} (2012)
  034}, [\href{http://arxiv.org/abs/0904.2715}{{\tt 0904.2715}}].

\bibitem{Tachikawa:2015bga}
Y.~Tachikawa, \emph{{A review of the $T_N$ theory and its cousins}},
  \href{http://dx.doi.org/10.1093/ptep/ptv098}{\emph{PTEP} {\bf 2015} (2015)
  11B102}, [\href{http://arxiv.org/abs/1504.01481}{{\tt 1504.01481}}].

\bibitem{Bergman:2014kza}
O.~Bergman and G.~Zafrir, \emph{{Lifting 4d dualities to 5d}},
  \href{http://dx.doi.org/10.1007/JHEP04(2015)141}{\emph{JHEP} {\bf 04} (2015)
  141}, [\href{http://arxiv.org/abs/1410.2806}{{\tt 1410.2806}}].

\bibitem{Hayashi:2014hfa}
H.~Hayashi, Y.~Tachikawa and K.~Yonekura, \emph{{Mass-deformed T$_{N}$ as a
  linear quiver}}, \href{http://dx.doi.org/10.1007/JHEP02(2015)089}{\emph{JHEP}
  {\bf 02} (2015) 089}, [\href{http://arxiv.org/abs/1410.6868}{{\tt
  1410.6868}}].

\bibitem{Kim:2012qf}
H.-C. Kim, J.~Kim and S.~Kim, \emph{{Instantons on the 5-sphere and
  M5-branes}},  \href{http://arxiv.org/abs/1211.0144}{{\tt 1211.0144}}.

\bibitem{Chang:2017cdx}
C.-M. Chang, M.~Fluder, Y.-H. Lin and Y.~Wang, \emph{{Spheres, Charges,
  Instantons, and Bootstrap: A Five-Dimensional Odyssey}},
  \href{http://dx.doi.org/10.1007/JHEP03(2018)123}{\emph{JHEP} {\bf 03} (2018)
  123}, [\href{http://arxiv.org/abs/1710.08418}{{\tt 1710.08418}}].

\bibitem{Chang:2017mxc}
C.-M. Chang, M.~Fluder, Y.-H. Lin and Y.~Wang, \emph{{Romans Supergravity from
  Five-Dimensional Holograms}},  \href{http://arxiv.org/abs/1712.10313}{{\tt
  1712.10313}}.

\bibitem{Herzog:2010hf}
C.~P. Herzog, I.~R. Klebanov, S.~S. Pufu and T.~Tesileanu, \emph{{Multi-Matrix
  Models and Tri-Sasaki Einstein Spaces}},
  \href{http://dx.doi.org/10.1103/PhysRevD.83.046001}{\emph{Phys. Rev.} {\bf
  D83} (2011) 046001}, [\href{http://arxiv.org/abs/1011.5487}{{\tt
  1011.5487}}].

\bibitem{deAlwis:1997gq}
S.~P. de~Alwis, \emph{{Coupling of branes and normalization of effective
  actions in string / M theory}},
  \href{http://dx.doi.org/10.1103/PhysRevD.56.7963}{\emph{Phys. Rev.} {\bf D56}
  (1997) 7963--7977}, [\href{http://arxiv.org/abs/hep-th/9705139}{{\tt
  hep-th/9705139}}].

\bibitem{Freedman:1998tz}
D.~Z. Freedman, S.~D. Mathur, A.~Matusis and L.~Rastelli, \emph{{Correlation
  functions in the CFT(d) / AdS(d+1) correspondence}},
  \href{http://dx.doi.org/10.1016/S0550-3213(99)00053-X}{\emph{Nucl. Phys.}
  {\bf B546} (1999) 96--118}, [\href{http://arxiv.org/abs/hep-th/9804058}{{\tt
  hep-th/9804058}}].

\bibitem{Penedones:2016voo}
J.~Penedones, \emph{{TASI lectures on AdS/CFT}},  in \emph{{Proceedings,
  Theoretical Advanced Study Institute in Elementary Particle Physics: New
  Frontiers in Fields and Strings (TASI 2015): Boulder, CO, USA, June 1-26,
  2015}}, pp.~75--136, 2017.
\newblock \href{http://arxiv.org/abs/1608.04948}{{\tt 1608.04948}}.
\newblock \href{http://dx.doi.org/10.1142/9789813149441_0002}{DOI}.

\bibitem{Klebanov:2011gs}
I.~R. Klebanov, S.~S. Pufu and B.~R. Safdi, \emph{{F-Theorem without
  Supersymmetry}}, \href{http://dx.doi.org/10.1007/JHEP10(2011)038}{\emph{JHEP}
  {\bf 10} (2011) 038}, [\href{http://arxiv.org/abs/1105.4598}{{\tt
  1105.4598}}].

\bibitem{Faddeev:1995nb}
L.~D. Faddeev, \emph{{Discrete Heisenberg-Weyl group and modular group}},
  \href{http://dx.doi.org/10.1007/BF01872779}{\emph{Lett. Math. Phys.} {\bf 34}
  (1995) 249--254}, [\href{http://arxiv.org/abs/hep-th/9504111}{{\tt
  hep-th/9504111}}].

\bibitem{Ruijsenaars:2000}
S.~N.~M. Ruijsenaars, \emph{On barnes' multiple zeta and gamma functions},
  {\emph{Advances in Mathematics} {\bf 156} (2000) 107--132}.

\bibitem{2003math......6164N}
A.~{Narukawa}, \emph{{The modular properties and the integral representations
  of the multiple elliptic gamma functions}}, {\emph{ArXiv Mathematics
  e-prints} (June, 2003) }, [\href{http://arxiv.org/abs/math/0306164}{{\tt
  math/0306164}}].

\bibitem{Faddeev:2012zu}
L.~D. Faddeev, \emph{{Volkov's Pentagon for the Modular Quantum Dilogarithm}},
  \href{http://dx.doi.org/10.1007/s10688-011-0031-8}{\emph{Funct. Anal. Appl.}
  {\bf 45} (2011) 291}, [\href{http://arxiv.org/abs/1201.6464}{{\tt
  1201.6464}}].

\end{thebibliography}\endgroup
\end{document}